# Security and Privacy Issues in Wireless Mesh Networks: A Survey

Jaydip Sen
Innovation Labs, Tata Consultancy Services Ltd. Kolkata, INDIA
email: jaydip.sen@acm.org

## 1. Introduction

Wireless mesh networking has emerged as a promising technology to meet the challenges of the next-generation wireless communication networks for providing flexible, adaptive, and reconfigurable architecture and offering cost-effective business solutions to the service providers [1]. The potential applications of *wireless mesh networks* (WMNs) are wide-ranging such as: backhaul connectivity for cellular radio access networks, high-speed *wireless metropolitan area networks* (WMANs), community networking, building automation, *intelligent transportation system* (ITS) networks, defense systems, and city-wide surveillance systems etc [2]. Although several architectures for WMNs have been proposed based on their applications [1], the most generic and widely accepted one is a three tier structure as depicted in Fig. 1. At the bottom tier of this architecture are the *mesh clients* (MCs) which are mobile devices (i.e. users) with limited mobility and having resource constraints in terms of power, memory and computing abilities. At the intermediate tier, a set of *mesh routers* (MRs) or *edge routers* form an interconnected wireless back bone – the *wireless mesh network* (WMN). The MRs are wireless routers which wirelessly connect with each other and provide connectivity to the MCs. At the top tier of the architecture are a group of *gateways* or *Internet gateways* (IGWs). Each IGW is connected with several MRs using wired links or high-speed wireless links. The IGWs are connected to the Internet by wired links. A mesh network thus can provide multi-hop communication paths between the wireless clients (i.e., the MCs), thereby serving as a community network, or can provide multi-hop connectivity between the clients and a gateway router (i.e. an IGW), thereby providing broadband Internet access to the clients. Since deployments of WMNs do not need any wired infrastructures, these networks provide a very cost-effective alternative to the *wireless local area networks* (WLANs) for the mobile users for the purpose of interconnection and access to the backbone Internet [2]. Wireless technology standards such as IEEE 802.11 (WLAN), IEEE 802.15 (LoWPAN), IEEE 802.16 (mobile WiMAX), IEEE 802.10 are adapted for developing a new wireless standard for mesh networking - IEEE 802.11s.

As WMNs become increasingly popular wireless networking technology for establishing the last-mile connectivity for home networking, community and neighbourhood networking, it is imperative to design efficient and secure communication protocols for these networks. However, the broadcast nature of transmissions in the wireless medium and the dependency on the intermediate nodes for multi-hop communications in such networks lead to several security vulnerabilities. These security loopholes can be exploited by potential external and internal attackers causing a detrimental effect on the network performance and disruption of services. The external attacks are launched by unauthorized users who intrude into the network for eavesdropping on the network packets and replay those packets at a later point of time to gain access to the network resources [3]. On the contrary, the internal attacks are strategized by some legitimate members in the network processing the authenticated credentials for accessing the network services. One example of such an attack is an intermediate node dropping packets which the node is supposed to forward. The internal attacks are more difficult to detect and prevent since the attackers are some members in the network having legitimate access to the network resources. Identifying and defending against these attacks in WMNs, therefore, is a critical requirement in order to provide sustained network services satisfying the quality of services of the user applications [4]. Furthermore, since in a WMN, the traffics from the end users are relayed via multiple wireless MRs, it is possible for these MRs to make a comprehensive traffic analysis for a user, thereby compromising the privacy his/her privacy. Hence, protecting the privacy and defending against privacy attacks on user data are critical requirements for most of the real-world applications in WMNs [5, 6]. Some security and privacy protection protocols for wireless networks are based on the computation and the use of the trust and reputation values of the nodes [7, 8]. However, most of these schemes are primarily designed for deployment in *mobile ad hoc networks* (MANETs) [9, 10], and hence these mechanisms do not fit well into the network architecture and the requirements of the applications in WMNs.

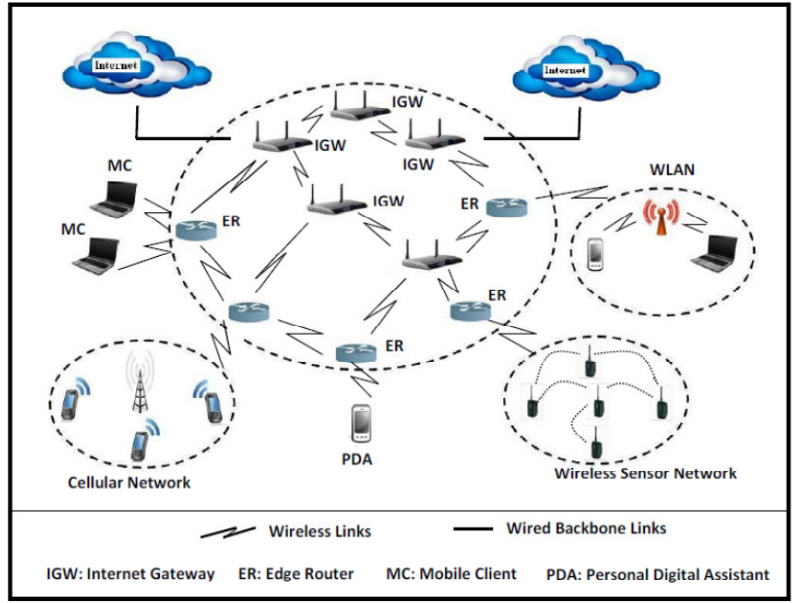

**Fig 1. The three-tier architecture of a wireless mesh network (WMN)**

Keeping in mind the critical requirement of security and user privacy in WMNs, this chapter provides a comprehensive overview of various possible attacks on different layers of the communication protocol stack for WMNs and their corresponding defence mechanisms. First, it identifies the security vulnerabilities in the physical, link, network, transport, application layers. Furthermore, various possible attacks on the key management protocols, user authentication and access control protocols, and user privacy preservation protocols are presented. After enumerating various possible attacks, the chapter provides a detailed discussion on various existing security mechanisms and protocols to defend against and wherever possible prevent the possible attacks. Comparative analyses are also presented on the security schemes with regards to the cryptographic schemes used, key management strategies deployed, use of any trusted third party, computation and communication overhead involved etc. The chapter then presents a brief discussion on various trust management approaches for WMNs since trust and reputation-based schemes are increasingly becoming popular for enforcing security in wireless networks. A number of open problems in security and privacy issues for WMNs are subsequently discussed before the chapter is finally concluded.

The chapter is organized as follows. Section 2 presents various possible attacks on different layers on the communication protocol stack of the WMNs. Section 3 discusses various security mechanisms at different layers for defending the attacks mentioned in Section 2. Section 4 provides a brief discussion on various trust management schemes for enforcing security and collaboration among the nodes in wireless networks with particular focus on WMNs. Section 5 highlights some future research trends on security and privacy issues in WMNs. Finally, Section 6 concludes the chapter.

**2. Security Vulnerabilities in WMNs**

Different protocols for various layers of WMN communication stack have several vulnerabilities. These vulnerabilities can be exploited by potential attackers to degrade or disrupt the network services. Since many of the protocols assume a pre-existing cooperative relation among the nodes, for successful working of these protocols, the participating nodes need to be honest and well-behaving with no malicious or dishonest intentions. In practice, however, some nodes may behave in a malicious or selfish manner or may be compromised by some other malicious users. The assumption of pre-existing trust relationships among the nodes, and the absence of a central administrator make the protocols at the link, network and transport layers vulnerable to various types of attacks. Furthermore, the application layer protocols can be attacked by worms, viruses, malwares etc. Various possible attacks may also be launched on the protocols used for authentication, key management, and user privacy protection. In this section, we present a comprehensive discussion on various types of attacks in different layers of WMN protocol stack.

## 2.1 Security vulnerabilities in the physical layer

The physical layer is responsible for frequency selection, carrier frequency generation, signal detection, modulation, and data encryption. As with any radio-based medium, the possibility of a jamming attack in WMNs is high since this attack can be launched without much effort and sophistication. Jamming is a type of attack which interferes with the radio frequencies that the nodes use in a WMN for communication [11]. A jamming source may be powerful enough to disrupt communication in the entire network. Even with less powerful jamming sources, an adversary can potentially disrupt communication in the entire network by strategically distributing the jamming sources. An intermittent jamming source may also prove detrimental as some communications in WMNs may be time-sensitive. Jamming attacks can be more complex to detect if the attacking devices do not obey the MAC layer protocols [12].

## 2.2 Security vulnerabilities in the link layer

Different types of attacks are possible in the link layer of a WMN. Some of the major attacks at this layer are: passive eavesdropping, jamming, MAC address spoofing, replay, unfairness in allocation, pre-computation and partial matching etc. These attacks are briefly described in this sub-section.

**(i) Passive eavesdropping:** the broadcast nature of transmission of the wireless networks makes these networks prone to passive eavesdropping by the external attackers within the transmission range of the communicating nodes. Multi-hop wireless networks like WMNs are also prone to internal eavesdropping by the intermediate hops, wherein a malicious intermediate node may keep the copy of all the data that it forwards without the knowledge of any other nodes in the network. Although passive eavesdropping does not affect the network, functionality directly, it leads to the compromise in data confidentiality and data integrity. Data encryption is generally employed using strong encryption keys to protect the confidentiality and integrity of data.

**(ii) Link layer jamming attack:** link layers attacks are more complex compared to blind physical layer jamming attacks. Rather than transmitting random bits constantly, the attacker may transmit regular MAC frame headers (with no payload) on the transmission channel which conforms to the MAC protocol being used in the victim network [13]. Consequently, the legitimate nodes always find the channel busy and back off for a random period of time before sensing the channel again. This leads to the denial-of-service for the legitimate nodes and also enables the jamming node to conserve its energy. In addition to the MAC layer, jamming can also be used to exploit the network and transport layer protocols [14]. Intelligent jamming is not a purely transmit activity. Sophisticated sensors are deployed, which detect and identify victim network activity, with a particular focus on the semantics of higher-layer protocols (e.g., AODV and TCP). Based on the observations of the sensors, the attackers can exploit the predictable timing behavior exhibited by higher-layer protocols and use offline analysis of packet sequences to maximize the potential gain for the jammer. These attacks can be effective even if encryption techniques such as *wired equivalent privacy* (WEP) and *WiFi protocol access* (WPA) have been employed. This is because the sensor that assists the jammer can still monitor the packet size, timing, and sequence to guide the jammer. Because these attacks are based on carefully exploiting protocol patterns and consistencies across size, timing and sequence, preventing them will require modifications to the protocol semantics so that these consistencies are removed wherever possible.

**(iii) Intentional collision of frames:** a collision occurs when two nodes attempt to transmit on the same frequency simultaneously [15]. When frames collide, they are discarded and need to be retransmitted. An adversary may strategically cause collisions in specific packets such as *acknowledgment* (ACK) control messages. A possible result of such collision is the costly exponential back-off. The adversary may simply violate the communication protocol and continuously transmit messages in an attempt to generate collisions. Repeated collisions can also be used by an attacker to cause resource exhaustion. For example a naïve MAC layer implementation may continuously attempt to retransmit the corrupted packets. Unless these retransmissions are detected early, the energy levels of the nodes would be exhausted quickly. An attacker may cause unfairness by intermittently using the MAC layer attacks. In this case, the adversary causes degradation of real-time applications running on other nodes by intermittently disrupting their frame transmissions.

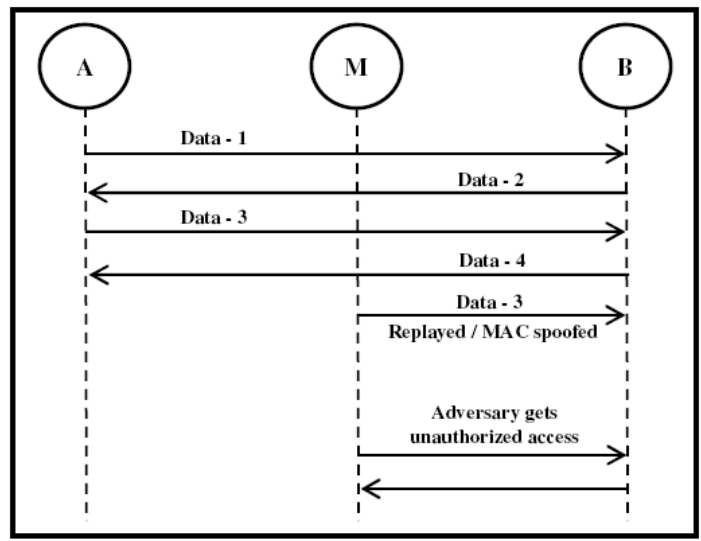

**Fig 2. Illustration of MAC spoofing and replay attacks launched by the malicious node *M***

**(iv) MAC spoofing attack:** MAC addresses have long been used as the singularly unique layer-2 network identifiers in both wired and wireless LANs. MAC addresses which are globally unique have often been used as an authentication factor or as a unique identifier for granting varying levels of network privileges to a user. This is particularly common in 802.11 WiFi networks. However, the MAC protocol in 802.11 standard and the network interface cards do not provide any safeguards against a potential attacker from modifying the source MAC address in its transmitted frames. On the contrary, there is often full support in the form of drivers from the manufacturers to change the MAC address in the transmitted frames. Modifying the MAC addresses in transmitted frames is referred to as *MAC spoofing*, and it can be used by attackers in a variety of ways. MAC spoofing enables the attacker to evade *intrusion detection systems* (IDSs) in the networks. Further, the network administrators often use MAC addresses in access control lists. For example, only registered MAC addresses are allowed to connect to the access points. An attacker can easily eavesdrop on the network to determine the MAC addresses of legitimate devices. This enables the attacker to masquerade as a legitimate user and gain access to the network. An attacker can even inject a large number of bogus frames into the network to deplete the resources (in particular, bandwidth and energy), which may lead to denial of services for the legitimate nodes.

**(v) Replay attack:** the replay attack, often known as the *man-in-the-middle* attack [16], can be launched by external as well as internal nodes. As shown in Fig. 2, an external malicious node (*M*) can eavesdrop on the broadcast communication between two nodes *A* and *B*. It can then replay the

(eavesdropped) messages later to gain access to the network resources. Generally, the authentication information is replayed where the attacker *M* deceives a node (node *B* in Fig. 2) to believe that the attacker is a legitimate node (node *A* in Fig. 2). On a similar note, the malicious node *M*, which is an intermediate hop between two nodes *A* and *B*, can keep a copy of all relayed data. It can then retransmit this data later to gain an unauthorized access to the network resources.

**(vi) Pre-computation and partial matching attack:** unlike the above-mentioned attacks, where MAC protocol vulnerabilities are exploited, these attacks exploit the vulnerabilities in the security mechanisms that are employed to secure the MAC layer of the network. Pre-computation and partial matching attacks exploit the cryptographic primitives that are used at the MAC layer for secure communication. In a pre-computation attack or *time memory trade-off attack* (TMTO), the attacker computes a large amount of information (key, plaintext, and respective ciphertext) and stores that information before launching the attack. When the actual transmission starts, the attacker uses the pre-computed information to speed up the cryptanalysis process. TMTO attacks are highly effective against a large number of cryptographic solutions. On the other hand, in a partial matching attack, the attacker has access to some (cipher text, plaintext) pairs, which in turn decreases the encryption key strength, and improves the chances of success of the brute force mechanisms. Partial matching attacks exploit the weak implementations of encryption algorithms. For example, in the IEEE 802.11 standard for MAC layer security in wireless networks, the MAC address fields in the MAC header are used in the *message integrity code* (MIC). The MAC header is transmitted as plaintext while the MIC field is transmitted in the encrypted form. Partial knowledge of the plaintext (MAC address) and the cipher text (MIC) makes IEEE 802.11i vulnerable to partial matching attacks.

DoS attacks may also be launched by exploiting the security mechanisms. For example, the IEEE 802.11i standard for MAC layer security in wireless networks is prone to the sensor hijacking attack and the man-in-the-middle attack, exploiting the vulnerabilities in IEEE 802.1X, and DoS attack, exploiting vulnerabilities in the four-way handshake procedure in IEEEE 802.11i.

### 2.3 Security vulnerabilities in the network layer

The attacks on the network layer can be broadly divided into two types: *control packets attacks* and *data packets attacks*. Furthermore, both these attacks could be either active or passive in nature [17]. Control packets attacks generally target the routing functionality of the network layer. The objective of the attacker is to make routes unavailable or force the network to choose sub-optimal routes. On the other hand, the data packet attacks affect the packet forwarding functionality of the network. The objective of the attacker is to cause the denial of service for the legitimate user by making user data undeliverable or injecting malicious data into the network. We first consider the network layer control packets attacks, and then the network layer data packets attacks.

**(i) Attacks on the control packets:** *Rushing* attacks that target the on-demand routing protocols (e.g., AODV), were among the first exposed attacks identified by Hu et al. [18] on the network layer of multi-hop wireless networks. Rushing attacks exploit the route discovery mechanism of on-demand routing protocols. In these protocols, the node requiring a route to the destination floods the *route_request* (RREQ) message, which is identified by a sequence number. To limit the flooding, each node only forwards the first message that it receives and drops remaining messages with the same sequence number. The protocol specify a specific amount of delay between receiving the RREQ message by a particular node and forwarding it, to avoid collusion of these messages. The malicious node launching the rushing attack forwards the RREQ message to the target node before any other intermediate node from the source to destination. This can easily be achieved by ignoring the specified delay. Consequently, the route from the source to the destination includes the malicious node as an intermediate hop, which can then drop the packets of the flow resulting in data plane DoS attack.

Hu et al. identified the *wormhole* attack that has a similar objective as that of the rushing attack but it uses a different strategy [19]. During a wormhole attack, two or more malicious nodes collude together by establishing a tunnel using an efficient communication medium (i.e., wired connection or high-speed wireless connection etc.), as shown in Fig. 3. During the route discovery phase of the on-demand routing protocols, the RREQ messages are forwarded between the malicious nodes using the established tunnel. Therefore, the first RREQ message that reaches the destination node is the one forwarded by the malicious nodes. Consequently, the malicious nodes are added in the path from the source to the destination. Once the malicious nodes are included in the routing path, the malicious nodes either drop all the packets, resulting in complete denial of service, or drop the packets selectively to avoid detection.

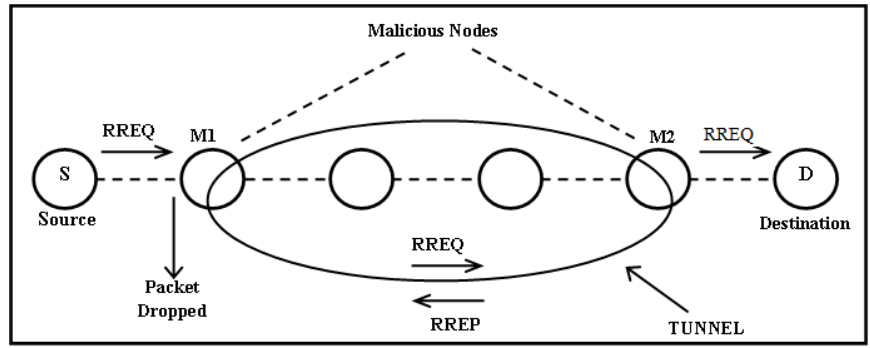

**Fig 3. Illustration of wormhole attack launched by nodes *M1* and *M2***

A *blackhole* attack (or *sinkhole* attack) [20] is another attack that leads to denial of service in WMNs. It also exploits the route discovery mechanism of on-demand routing protocols. In a blackhole attack, the malicious node always replies positively to a RREQ, although it may not have a valid route to the destination. Because the malicious node does not check its routing entries, it will always be the first to reply to the RREQ message. Therefore, almost all the traffic within the neighborhood of the malicious node will be directed towards the malicious node, which may drop all the packets, causing a denial of service. Fig. 4 shows the effect of a blackhole attack in the neighborhood of the malicious node where the traffic is directed towards the malicious node. A more complex form of the attack is the cooperative blackhole attack where multiple nodes collude together, resulting in complete disruption of routing and packet forwarding functionality of the network. Ramaswamy et al. have proposed a scheme for prevention of cooperative blackhole attack in which multiple blackhole nodes cooperate to launch a packet dropping attack in a wireless ad hoc network [21].

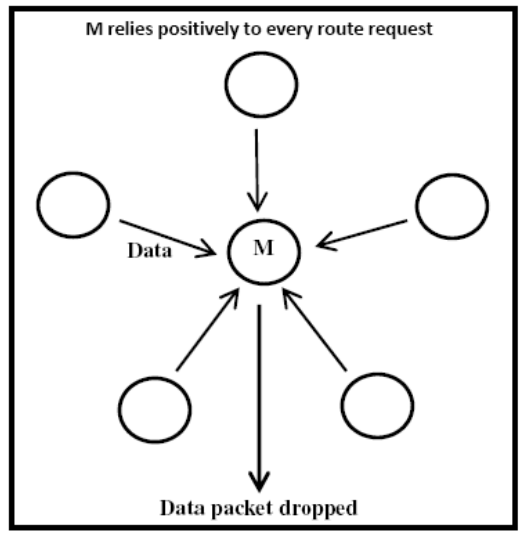

**Fig 4. Illustration of blackhole attack launched by node *M***

A *grayhole* attack is a variant of the blackhole attack. In a blackhole attack, the malicious node drops all the traffic that it is supposed to forward. This may lead to possible detection of the malicious node. In a grayhole attack, the adversary avoids the detection by dropping the packets selectively. A grayhole does not lead to complete denial of service, but it may go undetected for a longer duration of time. This is because the malicious packet dropping may be considered congestion in the network, which also leads to selective packet loss. Sen et al. have proposed a cooperative detection scheme for grayhole attack in a wireless ad hoc network [22].

A *Sybil* attack is the form of attack where a malicious node creates multiple identities in the network, each appearing as a legitimate node [23]. A Sybil attack was first exposed in distributed computing applications where the redundancy in the system was exploited by creating multiple identities and controlling the considerable system resources. In the networking scenario, a number of services like packet forwarding, routing, and collaborative security mechanisms can be disrupted by the adversary using a Sybil attack. Following form of the attack affects the network layer of WMNs, which are supposed to take advantage of the path diversity in the network to increase the available bandwidth and reliability. If the malicious node creates multiple identities in the network, the legitimate nodes, assuming these identities to be distinct network nodes, will add these identities in the list of distinct paths available to a particular destination. When the packets are forwarded to these fake nodes, the malicious node that created the identities processes these packets. Consequently, all the distinct routing paths will pass through the malicious node. The malicious node may then launch any of the above-mentioned attacks. Even if no other attack is launched, the advantage of path diversity is diminished, resulting in degraded performance.

In addition to the above-mentioned attacks, the network layer of WMNs are also prone to various types of attack such as: *route request (RREQ) flooding attack*, *route reply (RREP) loop attack*, *route re-direction attack*, *false route fabrication attack*, *network partitioning* attack etc. *RREQ flooding* is one of the simplest attacks that a malicious node can launch. An attacker tries to flood the entire network with the RREQ message. As a consequence, this causes a large number of unnecessary broadcast communications resulting in energy drains and bandwidth wastage in the network. A *routing loop* is a path that goes through the same nodes over and over again. As a result, this kind of attack will deplete the resources of every node in the loop.

Fig. 5 describes two instances where *route re-direction attack* has been launched by a malicious node *M*. In case *A*, the malicious node *M* tries to initiate the attack by modifying the mutable fields in the routing messages. These mutable fields include hop count, sequence numbers and other metric-related fields. The malicious node *M* could divert the traffic through itself by advertising a route to the destination with a larger *destination sequence number* (DSN) than the one it received from the destination. In case *B*, route re-direction attack may be launched by modifying the metric field in the AODV routing message, which is the hop-count field in this case. The malicious node *M* simply modifies the hop count field to zero in order to claim that it has a shorter path to the destination.

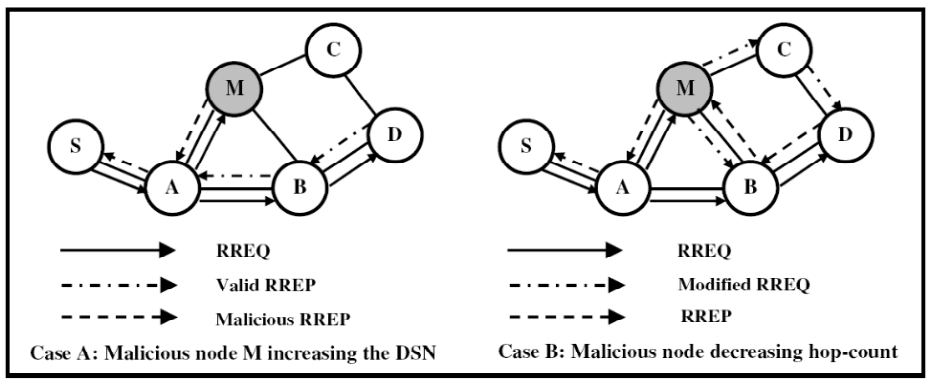

**Fig 5. Illustration of the route re-direction attack**

An adversary may fabricate false routing messages in order to disrupt routing in the network. For example, a malicious node may fabricate a *route error* (RERR) message in the AODV protocol. This may result in the upstream nodes re-initiating the route request to the unreachable destination so as to discover and establish alternative routes to them leading to energy and bandwidth wastage in the network. In a network partitioning attack, the malicious nodes collude together to disrupt the routing tables in such a way that the network is divided into non-connected partitions, resulting in denial of service for a certain network portion. Routing loop attacks affect the packet-forwarding capability of the network where the packets keep circulating in loop until they reach the maximum hop count, at which stage the packets are simply dropped.

**(ii) Attacks on the data packets:** the attacks on the data packets are primarily launched by selfish and malicious (i.e., compromised) nodes in the network and in the network and lead to performance degradation or denial of service of the legitimate user data traffic. The simplest of the data plane attacks is *passive eavesdropping*. Eavesdropping is a MAC layer attack. Selfish behavior of the participating WMN nodes is a major security issue because the WMN nodes are dependent on each other for data forwarding. The intermediate-hop selfish nodes may not perform the packet-forwarding functionality as per the protocol. The selfish node may drop all the data packets, resulting in complete denial of service, or it may drop the data packets selectively or randomly. It is hard to distinguish between such a selfish behavior and the link failure or network congestion. On the other hand, malicious intermediate-hop nodes may inject junk packets into the network. Considerable network resources (i.e., bandwidth and packet processing time) may be consumed to forward the junk packets, which may lead to denial of service for legitimate user traffic. The malicious nodes may also inject the maliciously crafted control packets, which may lead to the disruption of routing functionality. The control plane attacks are dependent on such maliciously crafted control packets. The malicious and selfish behaviors of nodes in WMNs have been studied in [24, 25]. The multi-hop wireless networks such as *mobile ad hoc networks* (MANETs), *wireless sensor networks* (WSNs), and *wireless mesh networks* (WMN) have many common security vulnerabilities in the network layer. Detailed discussions on various attacks on the network layer and their defense mechanisms for WSNs and WMNs can be found in [26] and [4] respectively.

**(iii) Attacks on multicast routing protocols:** multicast routing protocols deliver data from a source node to multiple destinations which are organized in a multicast group. Since many of the applications that use multicast services in a WMN have high-throughput requirements, and hop-count does not serve as a good metric for maximizing throughput, some protocols [27, 28] focus on maximizing path throughput, where paths are selected based on metrics that are dependent on the wireless link qualities. In these protocols, nodes periodically send probes to their neighbors to measure the quality of the links from their neighbors. Selection of the best path for maximizing throughput is done based on collaboration of nodes. An aggressive strategy for the best path selection assuming a perfect collaboration among all participating nodes provides an easy opportunity to a malicious node to manipulate the link metrics to its own advantage. In other words, an attacker may suitably adjust the link metrics so that it gets selected on the best routing path for a source-destination pair. In this way, it draws more traffic towards itself. However, since its intention is to disrupt network communication, it starts dropping packets which can lead to a possible network partitioning or can help the malicious node to carry out a traffic analysis on the network. Roy et al. have proposed a secure multicast routing protocol on a tree-based architecture of a WMN using hop-count as the metric for path selection [29]. In Section 3.3.11, we have discussed various attacks on the multicast routing protocols for wireless networks.

**2.4 Security vulnerabilities in the transport layer**

The attacks that can be launched on the transport layer of a WMN are: (i) *SYN flooding attack*, (ii) *de-synchronization attack*, and (iii) *session hijacking attack*.

*SYN flooding attacks* are easy to launch on a transport layer protocol like TCP. TCP requires state information to be maintained at both ends of a connection between two nodes, which makes the protocol vulnerable to memory exhaustion through flooding. An attacker may repeatedly make new connection request until the resources required by each connection are exhausted or reach a maximum limit. In either case, further legitimate requests will be ignored. One variant of such DoS attacks is the SYN flooding attack, in which an attacker creates a large number of half-open TCP connections with a target node without completing any of these requests. In the TCP protocol, two nodes have to successfully complete a *three-way handshake* mechanism before a session can be established between the pair of nodes. As shown in Fig. 6, in the first message, the node initiating the communication sends a SYN packet to the receiver node along with a sequence number. The receiver node sends a SYN/ACK message containing a sequence number and an acknowledgment sequence number to the initiator node. The initiator node then completes the handshake process by sending an ACK message containing an acknowledgment number. An attacker can exploit this protocol by sending a large number of SYN packets to a target node and spoofing the return address of the SYN packets. The SYN/ACK packets are sent back by the target node to the spoofed return address. The target node also waits for the final ACK message from the attacker keeping the half-open data structure open in its memory. When the number of such half-open connections becomes too high to create an overflow in the table which stores these data structures in the target victim node, the victim node will not be able to accept any further connections requests even from any legitimate nodes in the network, causing disruption in the network services.

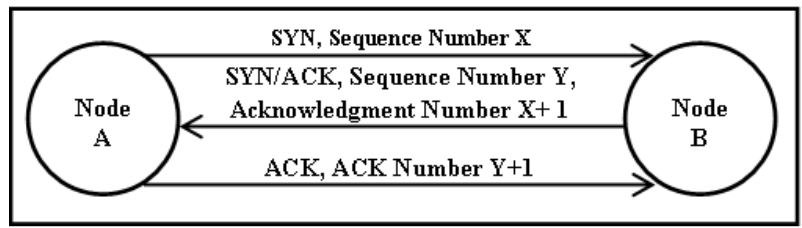

**Fig 6. Three-way handshake protocol for establishing a TCP session**

Session hijacking attacks exploits the vulnerability of the transport protocols (e.g., TCP) that do not provide any security check during an on-going session. All security mechanisms are applied only during the session establishment time. In a TCP session hijacking attack, an attacker spoofs the IP address of a victim node, correctly determines the current sequence number that is expected to be generated at the victim node, and then performs a DoS attack on the victim node.

A *de-synchronization attack* refers to the disruption of an existing connection [15]. An attacker may, for example, repeatedly spoof messages to an end host causing the host to request the retransmission of missed frames. If timed correctly, an attacker may degrade or even prevent the ability of the end hosts to successfully exchange data causing them instead to waste energy attempting to recover from errors which never really exist. Wu et al. have illustrated the de-synchronization attack that leads to *TCP ACK storm problem* [30]. In this attack, an attacker injects false data in an ongoing session between two nodes by launching a session hijacking attack. The false injected data is received by one of the nodes in the communicating pair and on receipt of the data, the node sends an ACK to the other node. Since the node at the other end was not expecting the sequence number of this ACK packet, the node tries to re-synchronize the session with its communicating peer. This cycle goes on indefinitely as the ACK packets travelling back and forth in the network causes an ACK storm.

**2.5 Security vulnerabilities in the application layer**

At the application layer, a compromise requires a full knowledge of the communicating applications (e.g., application layer formatting for traffic snooping) as well as compromising all the lower layers. The impact of such attacks can be extremely detrimental. For example, a *flooding attack* can affect the availability of the victim node as well as a large portion of the network. *Snooping attack* at the application layer can affect the integrity of the messages being communicated. However, snooping attack at the application has very low likelihood of success due to various defense mechanisms which are usually in place for protecting the lower layers. The use of encryption and authentication schemes at the higher layers also ensures that integrity of the messages is protected. The attacks in the application layer are mainly due to the viruses, malwares and worms or the repudiation attacks launched by insider nodes [30].

Mobile viruses and worms contain malicious codes which spread or replicate rapidly in a network and in the hosts and launch various types of attacks such as memory exhaustion, information leakage, phishing etc. Some types of Internet worms can scan the IP address of the nodes in a network and then send probe packets to critical UDP and TCP ports which are found in the port scanning process. The worms then attack the hosts using some application.

Repudiation attacks launched in the application layer cannot be detected or prevented by deploying firewalls at the network layer or by end-to-end encryption of traffic at the transport layer. An attacker getting an access to the information in network or in a host by sophisticated techniques can repudiate having conducting such an activity. Detection of such attacks needs sophisticated intrusion detection systems at multiple layers.

**2.6 Security vulnerabilities in the authentication protocols**

Several vulnerabilities exist in different authentication protocols used in WMNs. Notable among these attacks are: (i) unauthorized access, (ii) spoofing attack, (iii) *denial of service* (DoS) attack, and (iv) compromised or forged MRs.

In *unauthorized access* attack, a user who is not authorized to access a resource gets access to the network services by masquerading a legitimate user. The masquerader gains all the privileges of the legitimate node. Once an attacker is successful in launching such an attack, it becomes extremely difficult for a security mechanism to detect the attacker. *Spoofing* is the act of forging a legitimate MAC or IP address of a node. IP spoofing is quite common in multi-hop communication networks like WMNs. In IP spoofing attack, an adversary inserts a false source address or puts the address of a legitimate node on the packets forwarded by it. Using such a spoofed address, the malicious attacker can intercept a termination request and hijack a session. In MAC address spoofing, the attacker modifies the MAC address in the transmitted frames originating from a legitimate node. MAC address spoofing enables attackers to evade *intrusion detection systems* (IDSs) that may be place in different nodes in a WMN. In DoS attacks, a malicious attacker sends a flood of packets to an MR, thereby making a buffer overflow in the router (i.e. in an MR). In one variant of such an attack, a malicious node can send false termination messages on behalf of a legitimate MC, thereby preventing a legitimate user from accessing network services.

In compromised or forged MR attack, an attacker is able to compromise one or more MRs in a network by physical tampering or logical break-in. The adversary may also introduce rogue MRs to launch various types of attacks in a WMN. The fake or compromised MRs may be used to attack the wireless link, thereby implementing attacks such as: passive eavesdropping, jamming, relay and false message injection, traffic analysis etc. The attacker may also advertise itself as a genuine MR by forging duplicate beacons procured by eavesdropping on legitimate MRs in the network. When an MC receives these beacon messages, it assumes that it is within the radio coverage of a genuine MR, and initiates a registration procedure. The false MR now can extract the secret credentials of the MC and can launch a spoofing attack on the network. This attack is possible in protocols which require an MC to be authenticated by an MR and not the vice versa [31].

**2.7 Security vulnerabilities in the key management mechanisms**

Since the robustness and security of the cryptographic protocols used in WMNs are dependent on the strength of the keys used, key management is a very critical security function in WMNs. The functions of a key management protocol include: key generation, storage, distribution, updating, revocation and providing certificate services to the legitimate nodes in the network. Sophisticated attacks may be launched by malicious attackers to get access to the keys stored in a node or during the transit of the key from the key issuing server to the nodes in a WMN. For example, any key exchange protocol based on the *Diffie-Hellman* (DH) key exchange protocol [32] is vulnerable to the *man-in-the-middle* attack [33]. The key management protocols which are based on issuing of certificates to the network nodes by a trusted key distribution server or by a trusted third party are all vulnerable to DoS attacks.

**2.8 Security vulnerabilities in the user privacy protection mechanisms**

Protection of user privacy is an important issue in wireless network communication. However, ensuring privacy of the users is difficult to achieve even if the messages in the network are protected, as there are no security solutions or mechanisms which can guarantee that data is not revealed by the authorized parties themselves [34]. Communication privacy cannot be assured with message encryption since the attackers can still observe who is communicating with whom as well as the frequency and duration of each communication session. In addition, unauthorized parties can get access to the location information about the positions of different MCs by observing their communication and traffic patterns. Hence, there is a need to ensure location privacy in WMNs as well. In Section 3.8, we will see how privacy can be protected with respect to message contents, data traffic and location information.

Table 1 presents a summary of various types of vulnerabilities in different layers of the communication protocol stack of a WMN and their possible defense mechanisms. The details of the different defense mechanisms are discussed in Section 3.

**Table 1. Summary of different attacks on WMN protocol stack and their countermeasures**

| Layer | Attacks | Defense Mechanisms |
|---|---|---|
| Physical | Jamming | Spread-spectrum, priority messages, lower duty cycle, region mapping, mode change |
| MAC | Collision | Error-correction code |
| | Exhaustion | Rate limitation |
| | Unfairness | Small frames |
| Network | Spoofed routing information & selective forwarding | Egress filtering, authentication, monitoring |
| | Sinkhole | Redundancy checking |
| | Sybil | Authentication, monitoring, redundancy |
| | Wormhole | Authentication, probing |
| | Hello Flood | Authentication, packet leashes by using geographic and temporal information |
| | Ack. Flooding | Authentication, bi-directional link authentication verification |
| Transport | SYN Flooding De-synchronization | Client puzzles, SSL-TLS authentication, EAP |
| Application | Logic errors Buffer overflow | Application authentication Trusted computing, Antivirus |
| Privacy | Traffic analysis, Attack on data privacy and location privacy | Homomorphic encryption, Onion routing, schemes based on traffic entropy computation, group signature based anonymity schemes, use of pseudonyms. |

**3. Security Mechanisms against Various Attacks in WMNs**

In this section, we present a detailed discussion on the various security mechanisms for defending the attacks that we mentioned in the mentioned in Section 2. We provide description of various defense techniques at each layer of the protocol stack - physical, link, network, transport and application. In addition, some secure authentication mechanisms, user privacy protection schemes, and key management protocols are also discussed.

### 3.1 Security mechanisms for the physical layer

The jamming attack at the physical layer can be defended by employing different spread-spectrum technologies such as frequency hopping and code spreading [15]. In *frequency hopping spread spectrum* (FHSS) [35], signals are transmitted by rapidly switching a carrier signal among many frequency channels using a pseudo-random sequence which is known to both the transmitter and the receiver. Since it will be impossible for an attacker to predict the frequency selection sequence a priori, it will be difficult for him/her to jam the frequency being used at a given point of time. The interference is also minimized as the signal is spread over multiple frequencies.

In *direct sequence spread spectrum* (DSSS), each data bit in the original signal is represented by multiple bits in the transmitted signal using a spreading code. The spreading code spreads the signal over a wider frequency band which is directly in proportion to the number of bits being used. The receiver can use the spreading code with the signal to recover the original data.

Both FHSS and DSSS prohibit an attacker to intercept the radio signals. In order to successfully eavesdrop on the signal, the attacker must know the frequency band, the spreading code, and the modulation techniques being used. Spread spectrum technology also reduces the chance of interference from other radios and electromagnetic signals.

### 3.2 Security mechanisms for the link layer

Use of error-correcting codes is a common strategy for defending against *frame collision attack* [15]. However, these codes also add additional processing and communication overhead. Although it is reasonably easier to detect any malicious collision of frames, no comprehensive defense mechanism against such an attack is known to us today.

A strategy for defending against *energy exhaustion attack* is to apply a *rate limiting MAC admission control* mechanism. This will allow the network to ignore the requests that intend to exhaust the energy of a battery driven *mesh client* (MC) node. Use of *time division multiplexing* (TDM) can be another effective strategy in which each node is allotted a time-slot for transmission of its packets [15]. However, this mechanism is vulnerable to the frame collision attack, even when it can ensure that there is no possibility of an indefinite postponement of packet transmission in the back-off algorithm in the MAC layer.

The effect of unfairness caused by a malicious attacker can be partially eliminated by using small frames. Use of smaller packets reduces the time for the attacker to capture the channel making it harder for the attacker to launch an attack [15]. However, this technique often reduces the throughput in the network due to more control overhead. In addition, it is susceptible to further unfairness as the attacker may try to retransmit quickly instead of waiting for a random interval of time.

Various other security mechanisms [36, 37] have been proposed for multi-hop wireless networks that can be applied to WMNs possibly with slight modifications. All of these schemes are based on *data confidentiality service*, *data and header integrity services*, and *robust key management service* provided by the underlying cryptographic mechanisms. The data confidentiality service provides protection against the *passive eavesdropping attack*. Although, an eavesdropper can still intercept the encrypted message, he/she cannot decrypt it for extracting any information from the message since he/she does not have any access to the encryption key. The data and header integrity services provide protection against MAC spoofing attacks. The integrity verification algorithm at the receiver node will be able to detect any message with spoofed MAC address since the message will fail integrity verification test. Replay attacks in multi-hop wireless networks can be avoided by using per-packet authentication and integrity verification [36]. These approaches are based on using a fresh key for each packet which is synchronously computed by the sender and the receiver before the packet is sent by the sender node. Any replayed packet which is encrypted by an outdated key fails the integrity check at the receiver node due to key mismatch and automatically gets discarded. Use of a fresh key for each message also protects the data from pre-computation and partial matching attacks. Since the pre-computed information needs to be applied on every message in order to decrypt it, an attack becomes extremely costly [17].

In the following sub-sections, we discuss some of the existing security mechanisms for the link and the *medium access control* (MAC) layer of WMNs.

### 3.2.1 Application of synchronous dynamic encryption system in mobile wireless domains

Soliman and Omari propose a stream-cipher cryptosystem named *synchronous dynamic encryption system* (SDES) for wireless environment that is based on permutation vector generation [36]. The proposed light-weight cryptographic scheme has a high level of security. Specifically, the protocol is robust against (i) key compromise, (ii) biased bytes analysis (an attack, in which the attacker can analyze the byte distribution in the transmitted data to derive the key in a key-stream in a stream cipher), (iii) integrity violation. The number of key exchanges between the supplicants (SUP), the *access points* (AP) and the *authentication server* (AS) is kept at the minimum in order to reduce the communication overhead and the possible vulnerability during the key exchange process. The SUPs and the APs are always kept synchronized with the AS with respect to their shared encryption keys in such a way that it is impossible for a malicious intruder to get synchronized with the AS with the dynamically changing shared secret key. The node registration process is simple and it is carried out only once during the initial registration of the node with the AS. For ensuring security, use of two types of shared keys is proposed: (i) *secret authentication keys* (SAK) and (ii) *secret session keys* (SSK). The AS generates and transmits the initial SAK to each SUP and AP. For all subsequent mutual authentication processes with the AS, each SUP and AP uses its shared SAK. Once an SUP is initially authenticated by the AS, the AS forwards the SUP's SAK to the AP with which the SUP is associated. This reduces the delay in the authentication process. The SSK is generated per-session basis between the APs and the SUPs. The validity of an SSK is only during the session for which it is generated. For communication between two APs, the generation and distribution of the SSK is done by the AS. However, for secure communication between two SUPs, the AP associated with the source SUP generates and distributes the SSK to each SUPs. Both the keys (SAK and SSK) are used in the process of shuffling the *permutation vectors* (PVs) during the encryption process.

Since the protocol uses stream ciphers, the encryption and decryption processes are fairly simple and light-weight. For encryption, the source node carries out an XOR operation between the plaintext data and the corresponding PV to produce the ciphertext, and sends the ciphertext to the receiver node. The receiver node performs the decryption process by XORing the ciphertext with the same PV (generated at the receiver node). For the next cycle of encryption/decryption process, both the nodes synchronously generate a new PV based on their shared SAK and SSK.

Since the keys SAK and SSK serve as the seeds for generation of the stream of PV, the security of the protocol depends on the way these keys are generated and managed. The authors have proposed three modes for the generation of SAK/SSK, each mode providing different levels of security and involving different computing overhead. The three modes of operations are: (i) static shared keys, (ii) stream of shared keys, and (iii) dynamic stream of shared keys. In the first mode, the secret keys at both the communicating nodes are not changed. This makes the scheme vulnerable to cryptanalysis and successful key compromise attack. Since the permutation vectors may lead to the same stream of keys in successive cycles, it is easy to launch known plaintext-ciphertext pair attack. While this mode provides a very low level of security, it is computationally efficient since no key management is required. In the second mode, the shared keys are dynamically generated and changed after each encryption/decryption cycle. This makes the protocol secure against the known plaintext-ciphertext pair attack since it is not possible to make an easy cryptanalysis on the cipher. In addition, this mode is also secure against biased byte analysis. The additional overhead is also very low since it involves only an extra addition operation. However, in case of multiple simultaneous sessions between two nodes, due to use of the same key streams for all the sessions, breaking of one session will break all the sessions. This mode, therefore, fails to provide independent security to multiple simultaneous sessions. In the third mode, which provides the highest level of security, the data being transmitted is also used in the key generation process. Since the key generation process involves the data transmitted in the session, different sets of shared keys are generated for multiple simultaneous sessions, thereby eliminating the security loophole of the second mode. Another advantage of this approach is that data integrity guarantees that keys are not compromised during the transit. If the cipher is manipulated during the transit, it would break the synchronization of the shared keys at the two nodes. The additional overhead in this mode is due to two extra addition operations. The authors have provided detailed simulation results demonstrating the performance of the protocol.

### 3.2.2 A threshold and identity-based key management and authentication scheme

Deng et al. [38] propose a distributed key management and authentication approach in multi-hop wireless ad hoc network using the concepts of *identity-based authentication* [39, 40] and *threshold secret sharing* [41]. The scheme proposed by the authors follows a self-organized approach that does not assume any *a priori* trust association between the nodes or any centralized trusted entity in the network. This is in contrast to the traditional PKI-based authentication for key distribution and management, wherein a trusted server is deployed to generate, distribute and manage the keys.

The scheme assumes that each node in the network has an IP address or an identity, which is unique and remains unchanged throughout the lifetime of the node in the network. Each node discovers the identities of its one-hop neighbor by a neighbor discovery mechanism. The key generation process has two phases: (i) distributed key generation and (ii) identity-based authentication. The key generation phase is responsible for distributing the master key and the public/private key pair to each node in a distributed manner. The generated private keys are used for authentication. Authentication is realized by identity-based mechanism.

In the threshold cryptography-based solution proposed by the authors, the network has a public/private key pair, which is called the *master key*. The master key is used for key generation. The master public key (say, PK) is generated by the key generator and it is known to all the nodes in the network. The master private key (say, SK) is shared among the nodes in a threshold cryptographic manner. While no node can reconstruct the master private key (secret key) alone, any $k$ nodes among the total $n$ nodes in the network can jointly reconstruct the key. It is, however, infeasible even for $k$ -1 nodes to reconstruct the key by colluding among themselves. At the time of joining the networks, a node needs to acquire its private key corresponding to its identity by requesting the *private key generation* (PKG) service from at least $k$ neighbor nodes. The identity of the node is used as its public key. The authors have proposed the computation of the public key as $QID = H$ ($ID \parallel Expire\_time$), where $H(\ )$ is a hash function, *ID* stands for the identity of the node, and the *Expire_time* is a time stamp expressing the time of validity of the public key. When the public key of a node expires, the node contacts at least $k$ neighbors and presents its identity and requests for the PKG services. In the proposed scheme, since all the nodes have the master private key, any of them can act as the PKG node for any other node. Each of the $k$ PKG service nodes generates a secret share of the new private key and sends the same to the requesting node. In this way, any group of $k$ nodes can act as the PKG nodes for rest of the nodes such that a potential adversary who is able to compromise less than $k$ nodes cannot get access to a node's private key. The private key generation process is depicted in Fig. 7.

The scheme uses each node's identity as its public key. Since the identity of a node can be much shorter than a 1024 bit RSA public key, less communication and storage overhead is incurred in transmitting and storing the keys. The communication overhead incurred by the scheme is mainly due to the key generation process. In the network bootstrapping time, all the $n$ nodes have to participate in the generation of master key pair which induces large delay in set up. In addition, each node needs to broadcast a key generation request to its $k$ neighbors at the time of joining the network. In response, each PKG service node has to send its share of the generated private key. All these messages involve appreciable communication overhead. However, a trade-off can be made between the level of security and communication overhead in the scheme. A lower value of *k* will reduce the communication overhead while providing a lower level of security (since fewer nodes need to be compromised by an adversary to get access to the private key of a node). For higher level of security requirement, a larger value of *k* should be chosen. The authors have experimentally shown how the master key generation time varies with the size of the network and the effect of the value of the parameter *k* on PKG service time and the ratio of successful PKG service.

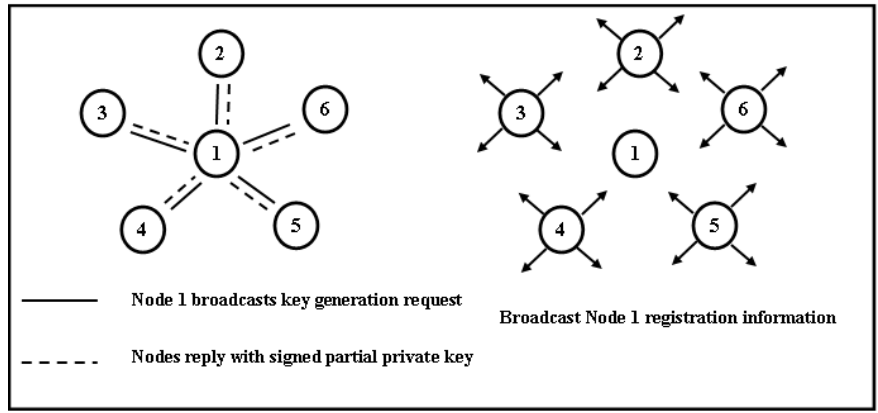

**Fig 7. The private key generation process of a node in Deng et al.'s scheme [38]**

### 3.2.3 Wireless intrusion detection and response mechanisms

Lim et al. [42] propose an intrusion detection system for wireless networks that consists of a number of devices deployed throughout the network. Each device is placed near an *access point* (AP) and all such devices are connected to a standard wired network to allow for remote management of the networked system. The intrusion detection system works at different levels. At the basic level, the system tracks the MAC address of the network adapter. If the MAC address is not found in the whitelist, or if it is found in the blacklist, then an alert is flagged. This is known as *MAC address filtering*.

The authors have also proposed to detect passive intruders using the IEEE 802.11b *request to send* (RTS) and *clear to send* (CTS) frames. The RTS frames are normally used to check whether the transmission medium is clear and to reserve a time slot for transmission of data. The CTS frames are used for acknowledging the RTS frames. The relationships between these frames may be used to detect presence of intruders in a network. If an active Wardriver is detected, RTS messages are sent to that MAC address. If the intruder is passively eavesdropping on the network, the card will respond with a CTS message, thereby revealing its presence. Stateful monitoring of packets in the network provides further detection of intrusions. Arrival of unexpected packets like unsolicited random responses might indicate a possible probing by an intruder.

In the proposed system, several detection devices are deployed that are connected to a central server so that it is possible to determine the exact position of an attacker or a rogue access point by *triangulation*. The position information may help in determining whether the source is a valid user with a possibly unregistered MAC address or a real intruder outside the premises. The central server may be augmented with additional authentication mechanisms such as *remote authentication dial-in user service* (RADIUS) authentication to actually identify whether a valid interface card is really being used by its assigned user or by some unauthorized person.

For intrusion response, the authors have suggested techniques like *address resolution protocol* (ARP) poisoning and disassociation-reassociation on the intruder. Since DoS attacks against the intruder will have an adverse impact on the overall network performance, a possible alternative is to send specially designed malformed frames targeted to the intruder. These frames may cause crashing on the intruder's computer. However, these intrusion response mechanisms are computationally expensive and their use will surely have an adverse impact on the network services.

### 3.2.4 MobiSEC: a security architecture for wireless mesh networks

Martignon et al. have presented a security architecture – MobiSEC – that provides access control in a WMN [43]. In this scheme, for authentication and key agreement between a node (an MC or an MR) with a *mesh access point* (MAP), a two-step approach is proposed. As shown in Fig. 8, in the first step, the new node (MC or MR) authenticates to the nearest MAP using 802.11i protocol [44]. In the second phase, the node uses a protocol based on *transport layer security* (TLS) and a certificate issued by a *certificate authority* (CA) with the AAA server to additionally authenticate as router and obtain the keying material required for this role in the WMN. For key distribution, use of two protocols is proposed – *server driven* and *client driven*. In the server driven protocol, each MR contacts a key distribution server for getting a key list. In the client driven protocol, the MRs obtains a seed from the server and a hash function type to generate the cryptographic keys as done in a hash chain method. Both the protocols need a mutual authentication based on certificate exchanges between the MRs and the key distribution server. MobiSEC supports mobility for both mesh clients and mesh routers. The client mobility is ensured since 802.11i protocol has client mobility support and MobiSEC is based on 802.11i authentication. The mobility of the routers in the backbone network is ensured by having all the routers using the same keying materials from the key server. Since all the routers in the backbone use the same key for authentication, router mobility in the backbone does not need any re-authentication process.

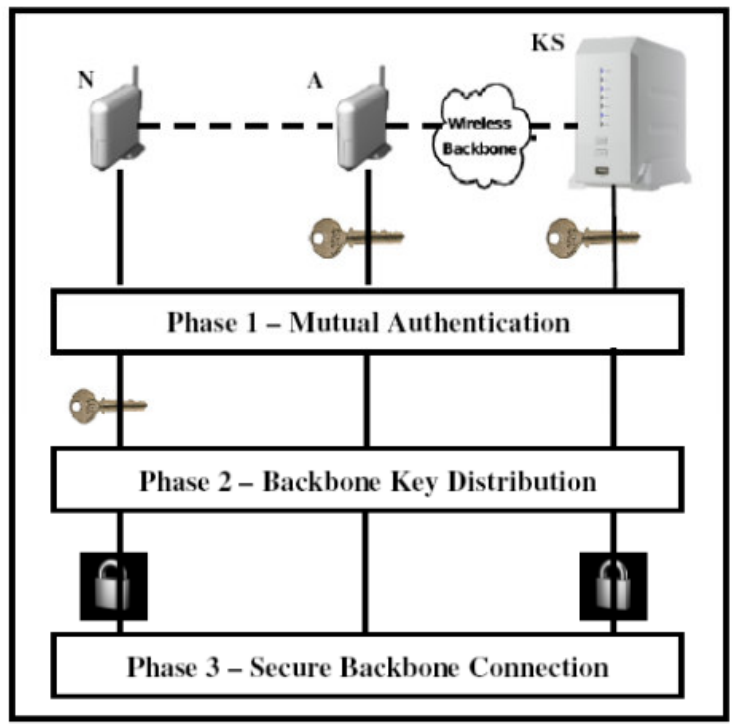

**Fig 8. Different phases of the connection process performed by a new mesh router *N* in MobiSEC**

The server driven protocol for key distribution is a reactive process for delivering the keys from the key distribution server to the mesh routers. This key is used for protecting the integrity and confidentiality of the traffic exchanged in the backbone during a specific interval. The protocol ensures that all the routers in the backbone have the same key for encryption and integrity protection of the messages transmitted in the mesh backbone network. In the client driven protocol, for key distribution, the key distribution server provides only a seed and a function type that should be used to compute the sequence of keys used by the mesh routers. The generation of the sequence of keys is similar to a hash chain computation, in which the computation of the next key is based on the output of a hash function to which the input was the key used in the previous round.

MobiSEC addresses access control issues including authentication and key establishment for the mesh clients and mesh routers in a WMN. However, the architecture does not explicitly addresses issues like message confidentiality, message integrity, and protection against replay attacks. In particular, the proposal only supports communications between the mesh clients and the mesh access points and between a pair of mesh routers [45]. In addition, use of a network-wide key for the protection of all messages in the mesh backbone is another issue which may lead to a complete breakdown of the security in backbone if a single mesh router is compromised. In addition, an attacker who is in possession of the backbone key can insert bogus traffic into the network thereby causing congestion and denial of service attack. Furthermore, the use of the mesh access point as the authenticator, implicitly assumes that key distribution server will transfer the keying material to the MAP during the authentication process. However, the mesh access point and the key server do have any shared secret for establishing a secure communication session between them, and only way to transfer the key material is to encrypt it using the key in the mesh backbone. If the backbone key is used for transferring the key from the key distribution server to the mesh access point, any malicious mesh router which is in neighborhood of the mesh access point will be able to capture the key. In spite of several security loopholes, MobiSEC provides a simple architecture for handling access control and mobility management issues in a WMN.

**3.2.5 Other security mechanisms for MAC layer misbehavior detection in WMNs**

Identifying various possible misbehaviors in the MAC layer and designing detection mechanisms for them has been a subject of extensive research in WLANs and ad hoc networks [46-48]. Some mechanisms for MAC layer misbehavior detection and their defense for WMNs have also been proposed [49-51].

Kyasanur and Vaidya have argued that the distributed contention resolution mechanism used in the MAC layer of IEEE 802.11 protocol is susceptible to abuse by a selfish node that does not adhere to the protocol and obtains an unfair share of the channel bandwidth [47]. To identify and penalize such selfish node, the authors have proposed a modification to 802.11 protocol. In the proposed modification, instead of the sender node selecting the random backoff time to initialize the backoff counter, the receiver node selects the backoff value and sends it in the *clear to send* (CTS) and ACK packets to the sender. The sender node uses this value of backoff in its next transmission to the receiver node. A receiver node can identify whether a sender node has deviated from the assigned backoff time by observing the number of idle slots between consecutive transmissions from the sender. If the observed number of idle slots is less than the assigned backoff, then there is a probability that the sender has deviated from the assigned backoff. The magnitudes of the observed deviations over a small number of packets transmissions are used to infer sender misbehavior with a high probability. If the sender node deviates from the assigned value, it will be assigned high backoff values in the next round to compensate for this deviation. However, this mechanism will be ineffective in case of a possible collusion between the sender and the receiver nodes or if the receiver node itself is a misbehaving node. Cardenas et al. have addressed the issue of preventing a possible colluding sender-receiver pair by ensuring randomness in the MAC protocol [52].

Konorski and Kurant have proposed a protocol called *R-hash* to prevent MAC layer misbehavior [53]. In the proposition, the winner of a contention is determined using a public hash function to the feedback each station gets from the contention. This confuses a potential misbehaving station is such a way that no modification of the probability distribution of transmission delay should be beneficial to these station.

Raya et al. have shown how a greedy user in a hotspot can substantially increase his/her share of bandwidth in the shared wireless medium by slightly modifying the driver of the network adapter of the wireless node [54]. A software system - DOMINO (Detection Of greedy behavior in the MAC layer of IEEE 802.11 public NetwOrks) - is designed that can detect and identify greedy stations without needing any modifications in the standard-compliant access points.

A proposition based on game theory for handling misbehavior in the MAC is been presented by Cagalj et al. [55]. The optimum strategy for each node has been derived in terms of controlling the channel access probability by adjusting the contention window, so that the equilibrium point is reached in the overall network. The authors have also derived conditions under which the Nash equilibrium of the network is Pareto optimal for each node in the network as well, when some of the nodes in the network are misbehaving.

Radosavac et al. have proposed a *minimax* robust MAC layer misbehavior detection framework, with the goal of having the optimum performance of the network under the worst-case attack scenario [46]. The network performance is measured using the required number of observations to arrive at a reliable decision. The framework not only captures the presence of an uncertainty in the attacks but also pays more attention to the attacks that are most significant in terms of their adverse impact on the network performance. It also considers scenarios in which an intelligent attacker launches an adaptive attack so that its detection becomes difficult.

Naveed and Kanhere have studied attacks on dynamic channel assignment in 802.11-based WMNs, in which a compromised mesh node manipulates control messages of the channel assignment protocol in such a way that the mesh links are forced to use heavily congested channels [51].

Table 2 presents a summary of the aforementioned MAC layer security schemes.

### 3.3 Security mechanisms for the network layer

A large number of schemes exist in the literature dealing with the issue of securing the network layer of WMNs [56-62]. In this section, we provide an overview of various security mechanisms in the network layer. A detailed discussion on these schemes can be found in [4].

As mentioned in Section 2.3, the attacks on the network layer can be either on the *route establishment* process or on the *data delivery* process, or on both. The protocols Ariadne [56] and SRP [63] intend to secure on-demand source routing protocols by using hop-by-hop authentication approach to prevent malicious packet manipulations in the route discovery process. On the other hand, SAODV [64], SEAD [57], and ARAN [58] use one-way hash chains to secure the propagation of hop counts in on-demand distance vector routing protocols. Papadimitratos and Hass have proposed a secure link state routing protocol that ensures correctness of the link state updates by using digital signatures and one-way hash chains [65]. To ensure correct data delivery, Marti et al. have presented two mechanisms - *watchdog* and *pathrater*- that can detect adversarial nodes by monitoring the packet forwarding behaviours of the nodes in a neighbourhood [59]. SMT [60] and Ariadne [56] use multi-hop routing to prevent malicious nodes from selectively dropping data packets. Sen et al. have proposed a co-operative detection scheme for identifying malicious packet dropping nodes in an ad hoc network that is robust in presence of Byzantine failure of nodes [66]. ODSBR protocol [61, 62] provides resilience to colluding Byzantine attacks by detecting malicious links based on end-to-end acknowledgment-based feedback technique. HWMP protocol [67, 68] allows two *mesh points* (MPs) to communicate using peer-to-peer paths. This model is primarily used if nodes experience a changing environment and no root MP is configured. While the proactive tree building mode is an efficient choice for nodes in a fixed network topology, HWMP does not address security issues and is vulnerable to a numerous attacks such as RREQ flooding attack, RREP routing loop attack, route re-direction attack, fabrication attack, tunnelling attack and so on [69]. LHAP [70] is a lightweight transparent authentication protocol for wireless ad hoc networks. It uses TESLA [71] to maintain the trust relationship among nodes.

In contrast to secure unicast routing, work studying security problems specific to multicast routing in wireless networks is particularly scarce. Two notable propositions on the secure multicast routing in wireless networks are [29] and [72]. Roy et al. propose an authentication framework [29] that prevents outsider attacks in a tree-based multicast protocol - MAODV [73]. Curtmola and Nita-Rotaru have presented a protocol named "BSMR" that addresses insider attacks in tree-based multicast protocols in wireless mesh networks [72].

**Table 2. Summary of some link and MAC layer defense mechanisms for WMN communication**

| Protocol | Salient Features |
|---|---|
| **SDES [36]** | It is a stream cipher-based cryptosystem for wireless networks that uses permutation vectors. The supplicants and the access points are always synchronized with the authentication server with respect to their shared keys so that it is impossible for an intruder to dynamically change the key and launch an attack. Use of stream ciphers makes the encryption and decryption processes fairly simple and light-weight. Two types of shared keys are used: (i) secret authentication keys (SAKs) and (ii) secret session keys (SSKs). Both these keys are used in the process of shuffling the permutation vectors during the encryption process. The protocol is robust against key compromise, biased bytes analysis, and integrity violation attacks. |
| **Threshold and identity-based key management [38]** | This authentication and key management scheme uses the concepts of identity-based authentication and threshold secret sharing. It assumes that each node has an IP address which is unique and remains unchanged throughout the lifetime of the network. The key generation process has two phases: (i) distributed key generation and (ii) identity-based authentication. In the key generation phase the master key and the public/private key pair are distributed to each node. The generated private key is used for authentication which is based on identity-based cryptography. The scheme is highly secure due to the deployment of a threshold authentication mechanism. |
| **Wireless intrusion detection and response system [42]** | The scheme proposes a wireless intrusion detection system (IDS) that consists of a number detection devices deployed in strategic points in a network. The IDS works at different level. At the basic level, it carries out a MAC address filtering if it cannot find the MAC address of a device in the white-list. For intrusion response, the system uses ARP poisoning and a disassociation-reassociation strategy with the suspected node. However, the proposed intrusion response mechanisms are computationally expensive and their invocation may adversely affect network performance. |
| **MobiSEC [43]** | It is an efficient scheme for secure authentication and access control in WMNs. It proposes a two-step approach for authentication of an MC with its MR. In the first step, the MC authenticates to the nearest MR. In the second phase, the MC uses a protocol that is based on the transport layer security and uses a certificate issued by a CA with the AAA server to additionally authenticate as a router. The key distribution may be server driven or client driven. In the server driven, each MR contacts a key distribution server for getting the key list, while in the client driven protocol, the MR obtains a seed from the server and a hash function to generate the key. The mobility of the MRs in the backbone is facilitated by having each router using the same key for authentication. The protocol addresses access control issues including authentication and key establishment. However, it does not address issues like message confidentiality, message integrity, and protection against replay attacks. |
| **R-hash [53]** | The scheme intends to prevent MAC layer misbehavior of nodes by using a hash function-based mechanism. The winner of a contention for accessing the wireless channel is determined by using a public hash function to the feedback that each station gets from the contention. This strategy effectively confuses a potential misbehaving station so that no possible modification can be made on the probability distribution of transmission delay for the contending stations. |
| **Game theory-based minimax framework [46]** | The goal of this game-theoretic proposition is to have a robust MAC layer misbehavior detection for optimizing the network performance under the worst-case attack scenario. It captures the presence of an uncertainty in the attacks and pays more attention to the attacks that are most significant in terms of their adverse impact on the network. The framework also considers adaptive strategy followed by sophisticated attackers which are very difficult to detect. |

A key point to note is that all of the above-mentioned secure protocols for unicast or multicast routing use only some basic routing metrics such as hop-count or latency. None of them consider routing protocols that incorporate high-throughput metrics, which are critical for achieving high performance in wireless networks. On the contrary, many of them even have to remove important performance optimizations in the existing protocols in order to prevent security attacks. There are also a few studies on secure QoS routing in wireless networks [74, 75]. However, these schemes are based on strong assumptions, such as existence of symmetric links, correct trust evaluation on nodes, ability to correctly determine link metrics even in an attack scenario etc. In addition, none of them consider attacks on the data delivery phase. Dong has proposed a scheme that considers both high performance and security as goals in multicast routing and deals with attacks on both path establishment and data delivery phases [76].

As mentioned in Section 2.3, wireless networks are also subject to attacks such as rushing attacks and wormhole attacks. Defences against these attacks have been extensively studied in [77-80]. RAP [18] prevents the rushing attack by waiting for several flood requests and then randomly selecting one to forward, rather than always forwarding only the first one. Techniques to defend against wormhole attacks include *packet leashes* [77] which restrict the maximum transmission distance by using time or location information, *Truelink* [79] which uses MAC level acknowledgments to infer whether a link exists or not between two nodes, and the use of directional antennas for detecting wormhole nodes [80].

In the following sub-sections, we provide brief discussions on some of the existing well-known secure routing protocols for WMNs. For more details on several such protocols, readers may refer to [4].

### 3.3.1 Authenticated routing for ad hoc networks (ARAN)

*Authenticated routing for ad hoc networks* (ARAN) is an on demand routing protocol that provides authentication of route discovery, route setup, and route path maintenance using cryptographic certificates [58]. It can detect and protect against malicious attackers without requiring any pre-deployed network infrastructure. However, it assumes a small amount of prior security coordination among the nodes. A trusted certificate server is used whose public key is assumed to be known to all nodes. On joining the network, each node receives a certificate issued by the trusted server. The certificate received by a node contains the IP address of the node, the public key of the node, the timestamp of creation of the certificate and the time at which the certificate would expire. A node uses its certificate for authenticating itself during the routing process. At the time of route discovery, a node broadcasts a signed *route discovery packet* (RDP). The RDP includes the IP address of the destination node, the certificate of the source node, a *nonce*, and a timestamp. The RDP is signed by the private key of the source node. Each node in the route discovery path validates the signature of the previous node, removes the certificate and the signature of the previous node, and records the IP address of the previous node. The node then signs the original contents of the packet, appends its own certificate and forwards the message after signing it with its private key. When the RDP reaches the intended destination node, the node creates a *route reply packet* (REP) and unicasts it back along the reverse path. The REP includes an identifier of the packet type, the IP address of the source, its certificate, the nonce, and the associated timestamp that was initially sent by the source node. On receiving the REP, the source node verifies the signature of the destination node, and the nonce. An *error message* (ERR) is generated if the timestamp or nonce does not match the requirements or if the certificate fails in the authenticity validation process. ARAN is a secure protocol that can prevent a number of attacks such as unauthorized participation of nodes, spoofed route signaling, spurious routing messages, alteration of routing packets, manipulation of the TTL values in the packets, and replay attacks. However, it is vulnerable to DoS attacks which are launched by flooding the network with bogus control packets. Since signature verification for each packet is required, the attacker can force a node to discard some of the control packets if the node cannot verify the signatures at the rate which is equal to or greater than the rate at which the attacker is injecting the bogus control packets.

### 3.3.2 Secure efficient ad hoc distance vector (SEAD) routing protocol

The *secure efficient ad hoc distance vector* (SEAD) [57] is a secure and proactive ad hoc routing protocol based on the *destination-sequenced distance vector* (DSDV) routing protocol [81]. The protocol deploys a one-way hash function for computing the hash chain elements which are used to authenticate the sequence numbers and the metrics of the update messages of the routing tables. The protocol ensures a mutual authentication between a source and a destination pair. The source of each routing table update message is also authenticated so as to prevent creation of any possible routing loop by an attacker which may try to launch an impersonation attack. Although the hash chains are useful for authenticating the metric and the sequence number, they are not sufficient for defending against a malicious node which can advertise the same distance and sequence number that the node has received. To defend against such malicious nodes, *hash tree chains* are used in conjunction with *packet leashes* [77], in which the address of the authenticator is tied with the address of the sender node. This prevents an attacker from replaying to an authenticator that it hears in its neighborhood. The protocol uses TESLA TIK [71] for shared key distribution among each pair of nodes in the network. SEAD can defend against routing loop attack if the loop does not contain more than one attacker. The protocol is simple and easy to implement by making a slight modifications to the DSDV protocol. The use of one-way hash chain for authentication reduces the computational complexity. The main drawback of the protocol, however, is the requirement of a trusted entity for distribution and maintenance of the verification element of each node. The trusted entity can also be a single-point-of-failure in the protocol operation.

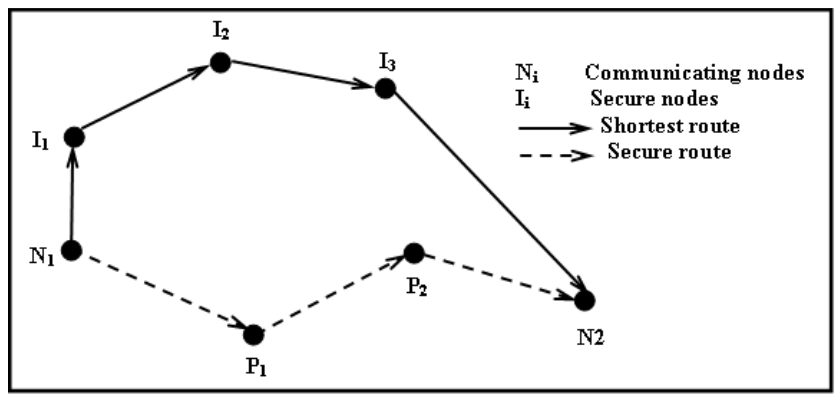

**Fig 9. Illustration of the use of trust metrics of nodes in SAR protocol**

### 3.3.3 Security-aware ad hoc routing (SAR) protocol

The *security-aware ad hoc routing* (SAR) protocol uses security as one of the key metrics in the route discovery and maintenance operations, and provides a framework for enforcing and measuring the attributes of the security metric [82]. Unlike traditional routing protocols which utilize distance (measured by the hop-counts), location, power and other metrics for routing path determination, SAR uses security attributes (such as trust values and trust relationships among nodes) in order to define a routing metric. SAR extends on-demand ad hoc routing protocols such as AODV [83] or DSR [84] in order to incorporate the security metric into the route discovery messages. The protocol ensures that an intermediate node that receives an RREQ packet can process or forward it only if the node can provide the required security or has the required authorization and trust level. If the node cannot provide the required security, the RREQ packet is dropped. If an end-to-end path with the required security attributes can be found, a suitably modified RREP message is sent from an intermediate node or the destination node. The security metric of SAR can be specified by a hierarchy of trust among the nodes. In order to define the trust levels, a key distribution or secret sharing mechanism is utilized in which the nodes belonging to a particular trust level share a key among them. Since the nodes of different security levels do not share any key, they cannot decrypt or process routing packets. SAR allows an application to choose its required level of security. However, the protocol needs different keys for different levels of security. Hence, with the increase in the number of security levels to be maintained, the number of keys to be managed also increases leading to an increase in storage and computational overheads.

Fig. 9 illustrates how trust metric is used in SAR. As shown in Fig. 9, the packets from the source node $N_1$ have two paths to travel to the destination node $N_2$. The shorter among these two paths, however, passes through nodes $P_1$ and $P_2$, whose trust levels are low. Hence, the protocol chooses a longer but secure path that passes through the trusted nodes $I_1$, $I_2$, and $I_3$.

### 3.3.4 Secure ad hoc on-demand distance vector (SAODV) routing protocol

The *secure ad hoc on-demand distance vector* (SAODV) routing protocol [64] is a secure extension of the AODV protocol [83]. The main objective of SAODV is to ensure integrity, authentication, and non-repudiation of the messages used in the AODV protocol. SAODV uses two mechanisms to secure routing messages: (i) digital signatures to authenticate the non-mutable fields of the messages, and (ii) hash chains to secure the hop count field which is the only mutable information in the packets. Since the protocol uses asymmetric cryptography for digital signatures, a key management mechanism is needed for enabling a node to acquire and verify the public key of other nodes in the network. SAODV uses the following additional fields in a routing packet header: (i) the *hash function* field identifies the one-way hash function used for securing the hop-count information, (ii) *max hop count* is a counter that specifies the maximum number of nodes a packet is allowed to go through, (iii) *top hash* field is the result of the application of the hash function on the max hop count times to a randomly generated number, and (iv) *hash* field is the random number used for routing. Each time a node sends an RREQ or an RREP message, it generates a random number and sets the value of the *max hop count* field same as the *time to live* (TTL) field in the IP header. The node then sets the *hash* field with the random number and also sets the *identifier* field of the hash function. Finally, the node computes the *top hash* by hashing the random number *max hop count* times. The protocol enables the receiver node to verify the hop count of each message by applying the hash function (*maximum hop count – hop count*) times to the value in the hash field. If the computed hash value and the value in the top hash field match, the hop count is successfully verified. Each time an RREQ message is re-broadcasted or an RREP is forwarded, the node has to apply the hash function to the hash field. Digital signatures are used to sign every field except the *hop count* and the *hash field*. Although the use of hash function and digital signature makes the scheme secure, the intermediate nodes cannot reply to an RREQ message if they have a fresh route to the destination node in their caches. In order to overcome this problem, the authors propose two solutions. The first solution does not allow the intermediate nodes to respond to a RREQ message and make then simply forward the RREQ message, since they cannot sign the message on behalf of the destination node. The second solution involves addition of a signature that can be used by intermediate nodes to reply to an RREQ by the node that originally created the RREQ. The *route error* (RERRs) messages are secured using digital signatures. A node that generates or forwards an RERR message, signs the whole message (except the destination sequence number) using its shared key with its neighbor node. Since the destination node does not authenticate the destination sequence number, a node should not update the destination sequence numbers of the entries in its routing table based on the RERR messages. The performance characteristics of SAODV are similar to those of the AODV protocol. However, the communication overhead in SAODV increases very rapidly with increase in mobility of the nodes due to the use of expensive asymmetric cryptographic operations.

### 3.3.5 Secure routing protocol (SRP)

The *secure routing protocol* (SRP) [63] is a secure extension that can be applied to many of the existing routing protocols especially to the DSR protocol [84]. The protocol requires the existence of a *security association* (SA) between a source-destination pair. This security association is utilized to establish a shared secret key between the two nodes. The protocol appends a header to each routing packet. The source node sends an RREQ with a *query sequence* (QSEQ) number which is used by the destination node to check whether the RREQ is outdated or valid, a random *query identifier* (QID) that identifies the specific request, and the output of a keyed hash function. The input to the function is the IP header, the header of the base protocol, and the shared secret key between the pair of nodes. The RREQ message generated by the source node is protected by a *message authentication code* (MAC) computed using the shared key between the source-destination pair. The RRQEs are broadcast to all the neighbors of the source node. Each neighbor that receives the RREQ for the first time appends its identifier to the RREQ and further broadcasts it in the network. All nodes maintain a priority ranking of its neighbors based on the rate at which the queries are generated from them. Higher priorities are assigned to nodes which generate queries at lower rates. The destination node checks the validity of the query and verifies its integrity and authenticity by computing and matching the keyed hash value. If the query is found to be valid and if it passes the integrity and authentication verification tests, the destination node generates a number of replies (RREPs) using different routes. This protects against attacks from malicious nodes that may attempt to modify the RREPs. An RREP includes the entire path from the source to the destination, the *query sequence* (QSEQ) number, and the *query identification* (QID) number. The integrity and authenticity of an RREP message is done using message authentication code in the same manner as in case of an RREQ message. Route maintenance is done using route error messages. The route error messages are source-routed along the path which is reported to be broken by an intermediate node. When the notified node receives a route error packet, it compares the route followed by the packet with the prefix of the corresponding route as reported in the route error packet. However, this approach has a security loophole since a fabricated route error attack can be easily launched by a malicious node. SRP is a light-weight protocol that can be easily implemented on a base routing protocol. However, as mentioned earlier, it cannot prevent unauthorized modifications of routes by malicious nodes.

**3.3.6 ARIADNE: a secure on-demand routing protocol for ad hoc networks**

Ariadne [56] is a secure on-demand routing protocol that is an extension of the *dynamic source routing* (DSR) protocol [84]. In contrast to the SEAD protocol [57] which is based on hop-by-hop authentication and message integrity, Ariadne assumes an end-to-end security approach. The protocol assumes the existence of a shared secret key between a pair of nodes and uses a *message authentication code* (MAC) for authenticating messages using this secret key. In fact, Ariadne proposes three schemes for authentication of messages: (i) authentication between two nodes using their shared secret key, (ii) shared secrets between communicating nodes combined with broadcast authentication using TESLA [71, 85], and (iii) digital signatures. In TESLA, a sender node generates a one-way key chain and defines a schedule based on which the keys are disclosed in the reverse order of their generation [71, 85]. This makes time synchronization a critical requirement for Ariadne. In the route discovery phase, the source node sends an RREQ message that includes the IP address of the source node, an ID that identifies the current route discovery process, a TESLA time interval for indicating the expected arrival time of the request to the destination, a hash chain that includes the address of the source node, the destination node address, the ID of the destination, and two empty lists – a *node list* and a *MAC list*. A neighbor, node on receiving the RREQ message, first checks the validity of the TESLA time interval so that the time interval is not too far in the future and its corresponding keys are not disclosed yet. A request with an invalid time interval is dropped by the neighbor nodes. If the time interval is valid, then the neighbor node inserts its address in the node list, replaces the hash chain with a new one that contains the address of the neighbor nodes along with the addresses of the nodes in the previous hash chain, and appends a *message authentication code* (MAC) of the entire packet to the MAC list. The MAC is computed using the TESLA key that corresponds to the time interval of the RREQ message. The neighbor node then broadcasts the RREQ message further in the network. The destination node buffers the RREQ and checks for its validity. An RREQ is considered to be valid if the keys with respect to the specified time interval have not yet been disclosed, and if the included hash chain can be verified. If the RREQ message is found to be valid, the destination node generates and broadcasts an RREP message. An RREP message contains all the fields of an RREQ message. In addition, it also contains a *target MAC* field and an empty *key list*. The target MAC field is filled in using the computed MAC of the preceding fields of the RREP message and the key that the destination shares with the initiator node. The RREP message is forwarded back to the initiator along the reverse path included in the node list as specified by the DSR protocol. An intermediate node, on receiving the RREP message, waits until the specified time interval allows it to disclose its key. On expiry of the specified time interval, the intermediate node discloses the key and appends the RREP to the key list and forwards the message to the next node. Upon receiving an RREP message, the initiator node verifies the validity of each key in the key list, checks the authenticity of the target MAC, and each MAC in the MAC list. The route maintenance in Ariadne is done in a similar manner as in DSR protocol. A node forwarding a packet to the next hop along the source route returns an RERR message to the packet's original sender if it is unable to deliver the packet to the next hop after a limited number of retransmission attempts. The most critical requirement for the operation of the Ariadne protocol is the existence of a clock synchronization mechanism. The base Ariadne protocol is vulnerable to wormhole attack. Hu et al. have proposed a security solution to defend against the wormhole attack using a mechanism called *packet leashes* [77].

### 3.3.7 Security enhanced AODV protocol

Li et al. have proposed a *security enhanced AODV* (SEAODV) routing protocol [69] that employs Bloom's key pre-distribution scheme [86] to compute *pair-wise transit key* (PTK) through the flooding of enhanced hello message. The protocol uses the established PTK to distribute the *group transit key* (GTK). The PTKs and GTKs are used for authenticating unicast and broadcast routing messages respectively. A unique PTK is shared between each pair of nodes, while the GTK is shared secretly between a node and all of its one-hop neighbors. A *message authentication code* (MAC) is attached as the extension to the original AODV routing message to guarantee the authenticity and integrity of the messages in a hop-by-hop manner. In order to ensure hop-by-hop authentication, each node must verify the incoming messages from its one-hop neighbors before re-broadcasting or unicasting the messages. The route discovery process in SEAODV is similar to that in AODV except for a minor difference. In SEAODV, an MAC extension is appended to the AODV routing packet. The MAC is computed based on the GTK of the node that broadcasts an RREQ message in its neighborhood. A neighbor node, on receiving the RREQ message, computes the MAC of the received message using the GTK. If the computed MAC matches with the received one, the received RREQ is considered to be authentic. The neighbor node then updates the hop-count of the RREQ message and its routing table. Further, it sets up a reverse path back to the source node by recording the node from which it has received the RREQ message. Finally, the node computes a message authentication code of the updated RREQ using the GTK and appends the MAC to the RREQ before re-broadcasting the RREQ. The destination node on receiving an RREQ generates an RREP message and unicasts it back to the source node along the reverse path. Since the RREP message is authenticated at each hop using the PTKs, an adversary can no way re-direct the traffic to some other route. A node generates a route error (RERR) message if it receives a packet for which it does not have an active route in its routing table, or the node possibly detects a broken link for the next hop of an active route. Although SEAODV is a secure extension of the AODV protocol, it is vulnerable to *RREQ flooding attack*. However, since the protocol provides authentication for RREQs from nodes that are in the list of active one-hop neighbors, such an attack would be detected very quite early before it can cause a serious damage in network communication.

### 3.3.8 Secure link state routing protocol (SLSP)

The *secure link state routing protocol* (SLSP) [65] is a secure proactive routing protocol for multi-hop wireless networks like MANET and WMNs. Its major goal is to enable a secure topology discovery and distribution of link state information across a wireless network. The critical requirement of SLSP protocol is the existence of an asymmetric key pair for every network interfaces of a node. The participating nodes in the network are identified by the IP addresses of their respective network interfaces. The key management is done by a group of nodes or by the use of *threshold cryptography* [41, 87]. The operation of SLSP can be logically divided into three parts: (i) public key distribution and management, (ii) neighbor discovery, and (iii) link state updates. The nodes broadcast their public key certificates within their zone using *public key distribution* (PKD) packets. The nodes verify the subsequent packets from the source node by matching its signed PKD packet. The link state information is also broadcasted periodically using *neighbor lookup protocol* (NLP) [65]. The NLP protocol uses signed *HELLO* messages which include the sender's MAC address and the IP address for the current network interface. NLP can inform SLSP about any suspicious observations (e.g. two different IP addresses having the same MAC address, or a node claiming the MAC address of the current node etc.) by generating notification messages. SLSP discards suspicious packets for which it has received a notification message. The hop count information in a packet is authenticated using hash chains. The *link state update* (LSU) packets are identified by the IP address of the initiating node [65]. The hash chains are authenticated using a digitally signed part of the LSU message. When a node receives an LSU it verifies the attached signature using a public key that it received earlier in the public key distribution phase of the protocol. To protect against DoS attacks, the nodes maintain a priority ranking of each neighboring node based on the rate of out-bound traffic. Nodes with lower rates of LSU generation are assigned higher priorities. This prevents a possible attack by a malicious node that may attempt to flood the network with spurious control packets, since the node will be always assigned a very low priority due its high rate of traffic generation. SLSP protocol provides security in the neighbor discovery process and uses NLP to identify spoofing attack by detecting discrepancies between the IP and the MAC addresses. However, the protocol is vulnerable to colluding malicious nodes that fabricate spurious links between themselves and flood this information in their neighborhood. Further, due to the use of asymmetric key cryptography, the protocol involves higher computational overhead.

### 3.3.9 Secure optimized link state routing (SOLSR) protocol

*Secure optimized link state routing* (SOLSR) protocol [88] is a secure extension of the base *optimized link state routing* (OLSR) protocol [89]. OLSR is a proactive link state routing protocol that employs an optimized flooding mechanism for diffusing link-state information. The optimization in OLSR is achieved by the use of *multi point relays* (MPRs). Fig.10 illustrates how the use of MPRs drastically reduces the overhead of control message communication.

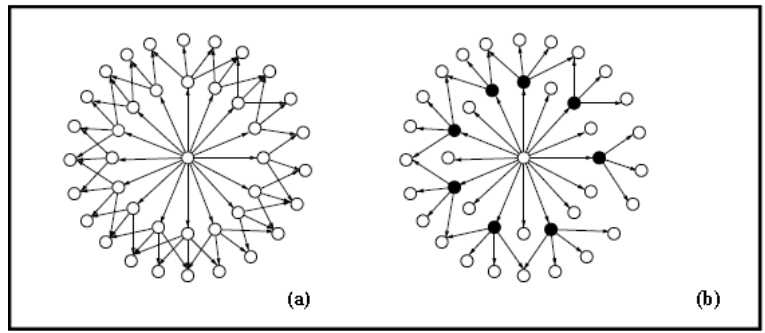

**Fig 10. OLSR: (a) Each 2-hop neighbor broadcasts. (b) Only MPRs transmit the broadcast [87]**

In OLSR, each node selects MPRs from among its neighbors in such a way that a message emitted by a node and further forwarded by the MPR nodes will be received by all nodes which are two-hops away from the source. Each node maintains its *MPR selector set*. On receiving an OLSR control message, a node consults its MPR selector set for deciding if the message is to be retransmitted. If the last hop of the control message is an *MPR selector*, then the message is to be retransmitted; otherwise, it is not retransmitted. If a message is to be broadcasted network-wide, it is sufficient to send it to a subset of the neighbors of the source node. This subset consists of the MPR set of the source node. In this way, OLSR optimizes message communication in a multi-hop wireless network. However, the OLSR protocol has a number of security vulnerabilities which can be exploited by a malicious node to launch attacks such as: (i) incorrect control traffic generation, (ii) incorrect *HELLO* message generation by identity spoofing or link spoofing, (iii) incorrect *topology control* (TC) message generation by identity spoofing or link spoofing, and (iv) incorrect control traffic relaying. The SOLSR protocol defends against such attacks by providing authentication for the OLSR signalling packets. The protocol uses *message authentication codes* (MACs) in every hop to ensure integrity and authentication of the routing messages. Every message is also time-stamped in order to ensure the freshness of the message. To prevent false message injection by malicious nodes, a signature is generated by the source node of each control message and the signature is appended with the control message. The receiver node checks the authenticity of the signature and the integrity of the message. Depending on the level of security desired, either an asymmetric key cryptographic method or a shared secret key cryptographic method is used for signature generation and verification and message integrity checking. The time stamps in the control messages are used to defend against replay attack. For each message that is generated by a source node, a unique timestamp is included in the message. If the difference between the time at which a message is received by a receiver and the timestamp of generation of the message at the source node does not exceed a threshold value than the message is considered to be fresh and it is accepted by the receiver provided it passes the authentication and integrity verification. There are various approaches for timestamp generation: (i) synchronous, real-time timestamps, (ii) non-volatile timestamps, and (iii) timestamps obtained using a challenge-response mechanism [88]. SOLSR involves more communication overhead than the base OLSR protocol. However, the computational overhead may be reduced by the use of symmetric key cryptography for message authentication and integrity verification purposes. The protocol is ideally suited to networks with low mobility like the WMNs. However, with a large network, it exhibits a scalability problem in its performance.

**3.3.10 Hybrid wireless mesh protocol (HWMP)**

Bahr has proposed a *hybrid wireless mesh protocol* (HWMP) [90]. It is the default routing protocol for IEEE 802.11s WLAN mesh networking. Every IEEE 802.11s compliant device is able to use this protocol for selecting routing paths. HWMP has both reactive and proactive routing capabilities. It is based on the adaptation of AODV routing protocol [83] into a novel protocol called r*adio-metric AODV* (RM-AODV) [91]. Unlike the AODV protocol that works on the network layer using the IP addresses, RM-AODV works on the MAC layer using the MAC addresses. RM-AODV uses a radio-aware metric for routing that helps in path selection. A *mesh portal* (a mesh point that has a connection to a wired network and acts as a bridge between the mesh network and the wired network) is configured to periodically broadcast mesh portal announcements to set up a tree with the mesh portal acting as the root of the tree. The created and maintained tree allows proactive routing with the mesh portal acting as the destination node. The proactive extension of HWMP uses the same distance vector routing strategy as RM-AODV and utilizes the routing control messages of RM-AODV for routing purpose. HWMP uses destination sequence numbers for detecting expired and outdated routing information. Routing packets with newer sequence numbers are always considered for routing and the packets with older sequence numbers are discarded. All routing table entries have specified validity time. The lifetime associated with a routing path is reset every time data frames are transmitted over that path.

The reactive components of HWMP uses a route discovery process which is similar to that used in the AODV [83] and the DSR [84] protocols. A source mesh points that needs to discover a path towards a destination mesh point broadcasts a route request (RREQ) packet. The destination mesh point or an intermediate mesh point that has a fresh route information to the destination node replies with a unicast route reply (RREP) message. However, the route discovery process in HWMP is adapted to the requirements of the IEEE 802.11s path selection protocol, and hence the MAC addresses of the nodes are used in routing and radio-aware links metrics are used for determining the optimal route path. The protocol uses the *airtime link metric* as defined by IEEE 802.11s standard [92] for this purpose.

HWMP has a proactive routing component as well. In deployment scenarios (for instance in a wireless mesh network that provides access to the Internet), large proportion of the traffic in a mesh network are destined for only one or a few mesh points. Since a proactive routing strategy to the mesh portal will be more efficient for such scenarios [90], the mesh portals are configured to periodically broadcast mesh portal announcements through wireless mesh network. A tree with the mesh portal as the root is constructed and a distance vector-based routing strategy as used in RM-AODV is adopted. The messages of RM-AODV are gainfully utilized in proactive routing.

The use of the proactive extension of RM-AODV and the reactive component of HWMP can be configured in the mesh portal node. This implies that the proactive component is optional in a mesh portal. For operation of the proactive component, a mesh portal is to be configured so that it can periodically broadcast mesh portal announcements. This triggers a root selection and routing tree construction process for the operation of the proactive routing protocol.

**3.3.11 Byzantine-resilient secure multicast routing (BSMR) protocol**

In multicast routing, data is delivered from a source node to multiple destination nodes which belong to a multicast group. Multicast routing protocols for wireless multi-hop networks use various approaches such as flooding, gossiping, geographical positions and are based on various communication structures such as meshes or trees. Designing a secure multicast routing protocol for wireless networks is more difficult than designing a unicast routing protocol due to several unique challenges that multicast communications bring in [72]. Curtmola and Nita-Rotaru have proposed a secure multicast routing protocol, named BSMR, that is resilient against Byzantine attacks [72]. The authors have first identified various possible attacks on multicast routing such as: Byzantine behavior of malicious nodes either alone or in collusion, which may lead to packet dropping, false packet injection, modification or replaying of packets etc at the network layer, intentional collision of frames at the MAC layer, and jamming at the physical layer. Further, in a multicast routing protocol, an adversary can attack the control messages for route discovery, route setup, and tree construction and management etc, and the data packets. In addition to attacks such as false route advertisement, generation of malicious route error messages may lead to network or multicast tree partitioning. Attacks on data packets include eavesdropping, modification, replay, false data injection, selective packet forwarding etc. Many of these attacks such as selective packet forwarding and DoS attacks cannot be prevented by use of cryptographic mechanisms only.

In the BSMR protocol [72], multicast data is communicated from the source to the members of a multicast group even if there are Byzantine attackers in the network as long as the multicast group members can be reached from the source node using paths that do not contain any adversarial node. An authentication mechanism is used that ensures that only authenticated nodes are allowed to perform certain critical operations such as joining in the multicast tree using valid group certificates. BSMR is also robust against a possible attack by a malicious node that may attempt to prevent a legitimate node from establishing a route to the multicast tree by flooding spurious route request or route reply messages. Selective packet forwarding attack is mitigated by using a reliability metric that detects adversarial behavior. The metric uses a list of link weights. A link with higher weight has lower reliability. Each node maintains its list corresponding to the weights of its links. This list is appended in each route request sent by the node so that the adversarial links are always avoided due to their higher weights when a new route to the tree is established. The reliability of a link is determined by the throughput of the link, and the nodes dynamically update their weight lists based on the link reliabilities. The authentication framework involves the use of a *tree token* by each of the authenticated members in the multicast tree. The tree token is periodically refreshed and distributed by the multicast group leader. The tree token is encrypted using the *pair-wise shared keys* established between each pair of neighbor nodes in the multicast tree. To allow any node in the network to verify whether the tree token possessed by a tree node is really a valid one, the group leader periodically broadcasts a *tree token authenticator*. The tree token authenticator can be expressed as *f (tree token)*, where *f* is a *collision-resistant one-way trap door function*. Any node can check the authenticity of a given tree token by applying the function *f* on it and checking the result with the value received from the tree token authenticator.

In order to prevent a node from falsely claiming that it is at a smaller hop distance from the group leader node than actually it is, the authors have proposed a technique based on *one-way hash chains*. The last element of the hash chain is referred to as the *hop count anchor*, which is periodically broadcasted in the network by the group leader thereby preventing a node to make any false claim about its distance from the group leader.

For joining a multicast group, a node needs to make a route discovery to the multicast tree. To prevent any possible attack, all route discovery messages are authenticated using the public key corresponding to the group certificate. All tree nodes use tree token to prove their membership in the current multicast group. For joining a multicast group, the requesting node first broadcasts an RREQ message that includes the node identifier, its weight list, the multicast group identifier, the last known group sequence number, and a request sequence number. The RREQ is flooded in the network till it reaches a tree node that has a group sequence number which is either greater than or equal to the group sequence number in the RREQ message. On receiving the RREQ message, the tree node initiates a response. The RREP message includes the node identifier, its recorded group sequence number, the requester's identifier, a response sequence number, the group identifier, and the weight list from the RREQ. To prove its current tree node status, the node also includes the current token encrypted with the requester's public key in the RREP message. The RREP is also flooded in the network till it reaches the requester node. The BSMR protocol uses a robust multicast tree maintenance strategy which is activated on occurrence of events such as pruning, link breaks, and network tree partitioning. The pruning messages are authenticated using the pair-wise keys shared between the tree neighbors. Even if a malicious node that has a sub-tree under it prunes itself, the legitimate nodes in the sub-tree will be able to reconnect to the tree using a procedure proposed in the protocol.

### 3.3.12 Secure on demand multicast routing protocol (SODMRP)

Dong et al. have proposed a secure version (SODMRP) [93] of the *on demand multicast routing protocol* (ODMRP) [94]. Before discussing the salient features of SODMRP, we first provide a brief overview of ODMRP.

ODMRP is an on-demand multicast routing protocol for multi-hop wireless networks. The protocol uses a mesh of nodes that constitutes a multicast group. Nodes are added to multicast groups using a route selection and activation protocol. The source node periodically reconstructs the mesh by flooding a JOIN QUERY message in the network so that membership information and the routing information are updated regularly. The interval between two successive mesh constructions is known as a *round*. JOIN QUERY messages are flooded in the network using a *basic flood suppression mechanism* which only allows the processing of the first received copy of a flooded message. When a JOIN QUERY message reaches a receiver node, the latter activates the path from itself to the source node by constructing a JOIN REPLY message and then broadcasting it. The JOIN REPLY message contains entries for each multicast group it wants to join. Each entry has a next hop field which is filled with the corresponding upstream node. When an intermediate node receives a JOIN REPLY message, it checks whether it is on the path to the source or not by verifying if the next hop field of any of the entries in the message matches with its own identifier. If the node finds that it lies on a path to the source, it makes itself a part of the mesh (the FORWARDING GROUP), and creates a new JOIN REPLY message using the matched entries. The node then broadcasts the JOIN REPLY message. As the JOIN REPLY messages reach the source node, the multicast receivers become connected to the source through a mesh of nodes (the FORWARDING GROUP) which guarantees delivery of multicast data. As long as a node is in the FORWARDING GROUP, it rebroadcasts any non-duplicate multicast data packets that it receives from its neighbors. To leave a multicast group, the receiver nodes just do not reply to the JOIN QUERY messages. They are not required to explicitly send any messages for this purpose. The participation of a node in the FORWARDING GROUP expires if its forwarding-node status is not updated in each time interval.

For enhancing the throughput of the ODMRP protocol Dong et al. first propose a high throughput algorithm called ODMRT-HT [93]. The fundamental differences between ODMRP and ODMRP-HT are: (i) unlike ODMRP which chooses links with minimum delay for routing, ODMRP-HT selects routes based on link quality metrics for achieving high throughput, and (ii) ODMRP-HT uses a *weighted flood suppression mechanism* to flood JOIN QUERY messages instead of a *basic flood suppression mechanism* [93]. Each node measures the link quality of each of the links with its neighbors based on a probing mechanism. The source node floods the JOIN QUERY message periodically which contains a route cost field based on the cumulative costs of the links of the route on which the message has travelled. When a node receives a JOIN QUERY message, it updates the route cost field by adding the metric of the last link over which the message has travelled. JOIN QUERY messages are flooded using a weighted flood suppression mechanism. In this approach, a node processes duplicate messages received over a fixed interval of time and rebroadcasts flood messages that advertise a better metric as indicated in the route cost field in the messages. Each node also records the node in the upstream path to the source node from which it has received the best link quality metric in the JOIN QUERY message. The receiver node, as in case of ODMRP, constructs a JOIN REPLY packet which is forwarded towards the source node through the *best path* as determined by the metric. The nodes on this best path are chosen as the members of the FORWARDING GROUP.

Dong et al. have identified various metric manipulation attacks that may be launched on ODMRP-HT protocol [93]. These attacks have been broadly categorized into two groups: (i) *local metric manipulation* (LMM) and (ii) *global manipulation* (GMM) [93]. Both these attacks types are Byzantine in nature since they may be launched by legitimate member nodes in the network which possess the necessary credentials. In the LMM attack, a malicious node intentionally increases the quality of its adjacent links and thereby creates a false perception among its neighbor about the link qualities. These falsely advertised good quality links have higher chances of being chosen by the neighbors and in this way the malicious node gets included on the selected routes. The GMM attack, on the other hand, involves a malicious node that arbitrarily changes the cumulative value of the route metric in a flooded packet before rebroadcasting it. In this way, the malicious node is able to not only manipulate its own contribution to the path metric in terms of its advertised link quality, but it can also adjust the contributions of the previous nodes on the routing path. Both these attacks are *epidemic* in nature and can have a detrimental effect on the performance of the throughput of the multicast routing.

To defend against the LMM and GMM attacks, Dong et al. have proposed the SODMRP protocol [93]. SODMRP uses an authentication framework which ensures that each node in the mesh network has a public-private key pair. In addition, each node possesses a *client certificate* that binds its public key to its one unique identity. Every packet is authenticated so that it is not possible for an outsider to inject any spurious packet in the network. For detection of attacks, two reactive approaches have been proposed: (i) a *measurement-based attack detection* protocol, and (ii) an *accusation-based reaction* protocol. The measurement-based attack detection strategy is based on the ability of the honest nodes in the network to detect discrepancy between the *expected packet delivery ratio* (ePDR) and the *perceived packet delivery ratio* (pPDR). The ePDR of a route is estimated from the value of the metric of the route, while the pPDR of a route can be determined by measuring the throughput along the route. Both the FORWARDING GROUP members and the received nodes monitor the pPDR along their routes. An alert is raised if the deference between the ePDR and pPDR exceeds a threshold value. In the accusation-based reaction, a node on detecting malicious behavior of another node, accuses the suspected node and floods the network with an ACCUSATION message. The ACCUSATION message contains the identity of the accuser as well as that of the accused node. The metrics advertised by the accused node are ignored and the accused node is not considered for inclusion in any subsequent FORWARDING GROUP selection. To prevent any possible *bad-mouthing attack*, a node is not allowed to issue any further accusation before the expiry of its previously made accusation. Any possible *metric poisoning effect* caused due to a metric manipulation attack is prevented by refreshing the metrics in the network immediately after an attack is detected. This is achieved by automatic and periodic broadcasting of JOIN QUERY messages.

SODMRP can defend against metric manipulation attack in wireless mesh networks and ensures high throughput in multicast communications. Since it uses asymmetric key cryptography, the computational overhead on the nodes and communication overhead in the network are higher which can be justified for applications which need high security and sustained high throughput.

Table 3 presents a summary of some of the aforementioned security schemes in the network layer of a WMN communication protocol stack.

**Table 3. Summary of some network layer security schemes for WMNs**

| Protocol | Salient Features |
|---|---|
| **ARIADNE [56]** | It is an on-demand routing protocol that assumes clock synchronization and the existence of a shared secret between each pair of nodes. It also assume an authentic TESLA key for each node in the network and an authentic route discovery chain element for each node for which this node will forward RREQs. TESLA keys are distributed to the participating nodes via an online key distribution center. Freedom from routing loop is guaranteed. Routing metric is the routing path length. In routing, shortest path identification is not done. Intermediate nodes are not allowed to reply to RREQs. It is resistant to: replay, DoS, routing table poisoning attacks. It is vulnerable to: location disclosure, black hole, wormhole attacks. |
| **SRP [63]** | It is an on-demand routing protocol that assumes the existence of a security association between each source and destination node. Malicious nodes are assumed not to collude. Freedom from routing loops guaranteed. Path length is the routing metric. The shortest path identification is not done. Intermediate nodes are allowed to optionally reply to RREQs. It is resistant to: replay, DoS, routing table poisoning attacks. It is vulnerable to: location disclosure, black hole, wormhole attacks. |
| **SAODV [64]** | It uses an on-demand routing approach that assumes the presence of an online key management scheme for association and verification of the public keys. Freedom from routing loops is guaranteed. Routing metric is the routing path length. It does not identify the shortest path in routing. Intermediate nodes are allowed to optionally reply to RREQs. It is resistant to replay, routing table poisoning attacks; and vulnerable to location disclosure, black hole, wormhole, DoS attacks. |
| **SEAD [57]** | It follows a table-driven (reactive) routing approach and assumes the existence of a clock synchronization, or a shared secret between each pair of nodes. Freedom from routing loop is guaranteed. Routing metric is the path length. It does not identify the shortest path in routing. Intermediate nodes are not allowed to reply to RREQs. It is resistant to replay, DoS, routing table poisoning attacks; vulnerable to location disclosure, blackhole, wormhole attacks. |
| **ARAN [58]** | It is an on-demand routing protocol that requires the presence of an online certification authority. Each node knows the public key of the CA a priori. Freedom from loops is guaranteed. Selection of the shortest path in routing is not mandatory. Intermediate nodes are not allowed to reply to RREQs. It is resistant to replay, routing table poisoning attacks; vulnerable to location disclosure, black hole, wormhole, DoS attacks. |
| **SMT [60]** | It is an on-demand routing protocol that assumes an initial trust between source and destination using public key cryptography. It also assumes a shared finite field for purposes of data dispersion in pre-computed set of columns. Freedom from routing loop is guaranteed. Selection of the shortest path in routing is not mandatory. Intermediate nodes may optionally reply to RREQs. It is resistant to replay, routing table poisoning attacks; vulnerable to location disclosure, black hole, wormhole, DoS attacks. |
| **SAR [82]** | This on-demand routing protocol assumes the existence of a key distribution or secret sharing mechanism. Freedom from routing loops not guaranteed- depends on the selected security requirements. Shortest routing path selection is not possible. Intermediate nodes are not allowed to reply to RREQs. It is resistant to replay, routing table poisoning attacks; vulnerable to location disclosure, black hole, wormhole, DoS attacks. |
| **SEAODV [69]** | It is an on-demand routing approach that assumes the presence of an online key management scheme for the association and verification of the public keys. Freedom from routing loops is guaranteed. Routing metric is the routing path length. It does not identify the shortest path in routing. Intermediate nodes are allowed to optionally reply to RREQs. It is resistant to replay, routing table poisoning attacks; vulnerable to location disclosure, blackhole, wormhole, DoS attacks. |
| **SLSP [65]** | It is a table-driven (proactive) protocol and assumes that the nodes have their public keys certified by a trusted third party (TTP). Malicious nodes are assumed not to collude. Freedom from loop is guaranteed. Routing metric is the routing path length. It does not involve any shortest path identification. Intermediate nodes are not allowed to reply to RREQs. It is resistant to replay, DoS, routing table poisoning attacks; vulnerable to: location disclosure, black hole, wormhole attacks. |
| **SOLSR [88]** | It is a table-driven (proactive) link state routing protocol and assumes a loose clock synchronization for time-stamping the messages. A key distribution center is also assumed to be present to manage the public keys or generation of the secret keys for message |

| | authentication, integrity and other security-related operations. Freedom from routing loop is guaranteed. Routing metric is the routing path length. It does not involve any shortest path identification. Intermediate nodes are not allowed to reply to the RREQs. It is resistant to replay attack, routing table poisoning attack, incorrect control traffic generation, incorrect HELLO message generation by identity spoofing or link spoofing, incorrect topology control (TC) message generation, incorrect control traffic relaying attacks; vulnerable to blackhole, wormhole, DoS and location disclosure attacks. |
|---|---|
| **SODMRP [93]** | It is an on-demand multicast routing protocol and assumes a public key cryptographic framework in place. It also assumes that each node possesses a client certificate that binds its public key to its unique identity. Freedom from routing loop is guaranteed. It ensures high throughput in multicast communications to support rich user experience. It is resistant to local metric manipulation (LMM) and global metric manipulation (GMM) attacks; vulnerable to minor bad mouthing (false accusation) by malicious nodes. |
| **TESLA [71, 85]** | It is a source authentication scheme for multicast communication which is based on loose time synchronization between the sender and the receivers followed by a delayed release of the authentication key by the sender. The sender attaches a MAC to each packet using the key which initially is known to the sender only; the receiver buffers it without being able to authenticate the packet. A short while late, the sender discloses the key and the receiver is then able to authenticate the packet. A single MAC per packet is able to ensure source authentication if the receiver has a synchronized clock which is ahead in time as that of the sender. TESLA is light-weight, scalable and can be used in the network or in the application layer. For security, the sender and the receiver must have a loose time synchronization in which the receiver needs to know an upper bound on its deviation from the sender' clock although precise time synchronization is not required. It is resistant to DoS attack on the sender if indirect time synchronization is used. However, it is prone to DoS attack on the sender if direct time synchronization is used. A powerful buffer overflow attack on the receivers and DoS attacks on the authentication key chains are also possible. |
| **Packet Leashes [77]** | It is a mechanism to defend against the wormhole attack in which an attacker records a packet at one location in a network and tunnels the data to another location and replays the packet there. The attacker can perform the attack even if cryptographic services providing confidentiality, authentication and integrity protection are available in the network. The scheme assumes that each node can obtain an authenticated key from any other node. A trusted entity is also assumed that signs the public-key certificates for each node. Two types of leashes are distinguished- geographical leashes and temporal leashes. In temporal leashes, an extremely precise clock synchronization mechanism is assumed to be present. In the geographical, the scheme assumes geographical location information and loosely synchronized clocks. Geographical leashes are less efficient, since they require broadcast authentication. The scheme is particularly designed to defend against wormhole attack to which most of the wireless routing protocols are vulnerable. |
| **HWMP [67,68]** | It is a hybrid routing protocol that has both reactive and proactive routing capabilities. It assumes the existence of a mesh portal that is configured to periodically broadcast beacons. Routing metric is the routing path length. The intermediate nodes can optionally respond to the RREQs. The route discovery process is adapted to the IEEE 80.11s path selection protocol in which MAC addresses of the nodes are used in routing. Optimal routing path is determined based on radio-aware link metrics. It is resistant to replay, routing table poisoning attacks; vulnerable to location disclosure, blackhole, wormhole, DoS attacks. |
| **Secure MAODV [29]** | It is a secure multicast on-demand routing protocol in which each node (multicast group members as well as non-members) possesses a pair of public/private keys and a certificate signed by a CA. The certificate binds the public key of a node to its IP address. A group member has a group membership certificate which binds the group member's public key and IP address with the IP address of the multicast group. Each node in a multicast tree establishes pairwise shared keys with its neighbor. The multicast group leader digitally signs the group HELLO packets to prevent spoofing. Tree key credentials are used for distinguishing between tree nodes and other nodes. The hop counts are authenticated using one-way has chains. It is resistant to outsider attacks and insider attacks such as attacks on route discovery process by a non-tree or a tree node, attacks on link activation, attacks on multicast tree maintenance – on the tree pruning process, on the link repair process, and on the partition merge process. It is vulnerable to DDoS attack. |
| **ODSBR [61, 62]** | It is an on-demand (reactive) routing protocol that assumes bi-directional communication links and requires pairwise shared keys among the nodes which are established on-demand. |

|  | The services of a public key infrastructure (PKI) are assumed to be available for key distribution, management and revocation. The protocol establishes a reliability metric based on the behavioral analysis of the nodes and selects the best path based on it. The metric is represented by a list of link weights. It is resistant to Byzantine attacks by insider nodes, such as creation of routing loops, routing packets via non-optimal paths, selectively dropping packets etc; vulnerable to DoS and wormhole attacks. |
|---|---|
| **BSMR [72]** | It is a multicast routing protocol that assumes a public key infrastructure to enable public key cryptographic operations. For multicast communication, a multicast tree is constructed. For joining a multicast group a new node makes a route discovery to the multicast tree by broadcasting RREQ messages. Tree maintenance algorithms are invoked on occurrence of events such as pruning, link breaks, and network tree partitioning. It is resistant to Byzantine attacks by insider nodes, packet dropping, false packet injection, modification or replaying of packets, false route advertisement, generation of malicious route error message that leads to network or multicast tree partitioning, intentional collision of frames at the MAC layer and jamming at the physical layer. However, it is vulnerable to DDoS attacks. |

### 3.4 Security mechanisms in the transport layer

*Secure socket layer* (SSL) [95], *transport layer security* (TLS) [95] and *private communications transport* (PCT) [95] protocols are usually used for securing the transport layer in wireless networks including the WMNs. SSL/TLS uses asymmetric key cryptographic techniques to ensure secure communication sessions. It can also help in protecting against masquerading attack, man-in-the middle attack, rollback attack, replay attack and buffer overflow attack.

For securing the transport layer in WMNs, an upper layer authentication protocol - *extensible authentication protocol encapsulating transport layer security* (EAP-TLS) protocol - is proposed by Aboba and Simon [96]. Although EAP-TLS offers mutual authentication between a *mesh router* (MR) and a *mesh client* (MC) or between a pair of MCs, it introduces high latency in WMNs because each terminal acts as an authenticator for its previous neighbor before the authentication request reaches an *authentication server* (AS). Furthermore, for nodes with high mobility, frequent re-authentications due to handoffs can have a very adverse impact on the quality of service of the applications. As a result, variants of EAP-TLS have been proposed to adapt IEEE 802.1X authentication model for multi-hop WMNs [3].

### 3.5 Security mechanisms in the application layer

The most usual ways of securing the application layers is the use of firewalls and *intrusion detection systems* (IDS). In a wireless network with a firewall installed, the firewall provides easy controls for achieving access control, user authentication, packet filtering, and logging and accounting services etc. Application-level firewalls provide protection against various attacks such as detection of malwares, spywares etc. However, the access control policies in a firewall are static in nature which makes a firewall unable to detect a new attack based on an anomaly-detection technique.

For detecting more sophisticated novel attacks, *intrusion detection systems* (IDSs) are used as the second line of defense along with firewalls. Interested readers may refer to [97, 98] for a comprehensive discussion on IDSs in wireless networks. An architecture of a cooperative, distributed IDS based on clustering of nodes for a multi-hop wireless network (CWIDS) that can detect attacks at multiple layers (including the application layer) is presented in [99]. An agent-based IDS for the network layer is proposed that can be adapted to large-scale distributed WMNs [100].

Table 4 presents a list of vulnerabilities in different layers of the protocol stack of WMNs and the corresponding security schemes for defending these attacks.

**Table 4. Summary of various attacks and their defense mechanisms for WMN communications**

| Attack | Targeted layer in the protocol stack | Protocols |
|---|---|---|
| Jamming | Physical and MAC layers | Frequency hopping spread spectrum (FHSS) [35], Direct sequence spread spectrum (DSSS) [35] |
| Wormhole | Network layer | Packet Leashes [77] |
| Blackhole | Network layer | SAR [82] |
| Grayhole | Network layer | GRAYSEC [22], SAR [82] |
| Sybil | Network layer | SYIBSEC [23] |
| Selective packet dropping | Network layer | SMT [60], ARIADNE [56], Sen [7], Sen[9], Sen [66] |
| Rushing | Network layer | ARAN [58], SAR [82], SEAD [57], ARIADNE [56], SAODV [64], SRP [63], SEAODV [69] |
| Byzantine | Network layer | ODSBR [61, 62] |
| Resource depletion | Network layer | SEAD [57] |
| Information disclosure | Network layer | SMT [60] |
| Location disclosure | Network layer | SRP [63] |
| Routing table modification | Network layer | ARAN [58], SAR [82], SRP [63], SEAD [57], ARIADNE [56], SAODV [64], SEAODV [69] |
| Multicast routing metrics manipulation attack | Network layer | SODMRP [93] |
| Byzantine multicast routing insider attack | Network layer | BSMR [72] |
| SYN flooding | Transport layer | SSL [95], TLS [95], EAP-TLS |
| Session hijacking | Transport layer | SSL [95], TLS [95], PCT [95], EAP-TLS[96] |
| Repudiation | Application layer | ARAN [58 ], CWIDS [99] |
| Denial of service | Multi-layer | SRP [63], SEAD [57], ARIADNE [56] |
| Impersonation | Multi-layer | ARAN [58], SEAD [57], SEAODV [69] |

### 3.6 Secure authentication mechanisms

Robust authentication and authorization mechanisms provide adequate safeguards against fraudulent access by unauthorized users in WMNs. Authentication ensures that an MC and the corresponding MR can mutually validate their credentials with each other before the MC is allowed to access the network services. Since secure authentication is a critical requirement for real-world deployments of WMNs, an extensive work has been done by researchers on this topic. In the following, we present a brief discussion on some of the secure authentication mechanisms and then provide a detailed discussion on four such propositions.

Mishra and Arbaugh propose a standard mechanism for client authentication and access control to guarantee a high-level of flexibility and transparency to all users in a wireless network [16]. The users can access the mesh network without requiring any change in their devices and softwares. However, client mobility can pose severe problems to the security architecture, especially when real-time traffic is transmitted. To cope with this problem, proactive key distribution approach is proposed [101,102].

Providing security in the backbone network for WMNs is another important challenge. Mesh networks typically employ resource constrained mobile clients. It is sometimes difficult to protect these devices against removal, tampering, or replication attacks. If a device can be remotely managed, a distant hacking into the device would work perfectly [103]. Accordingly, several research works have been done to investigate the use of cryptographic techniques to achieve secure communication in WMNs. Cheikhrouhou et al. have proposed a security architecture [104] that is suitable for multi-hop WMNs employing PANA (Protocol for carrying Authentication for Network Access) [105]. In the scheme proposed by the authors, the wireless clients are authenticated on production of the cryptographic credentials necessary to create an encrypted tunnel with the remote access router to which they are associated. Even though such framework protects the confidentiality of the information exchanged, it cannot prevent adversaries to perform active attacks against the network itself. For instance, a malicious adversary can replicate, modify and forge the topology information exchanged among mesh devices, in order to launch a denial of service attack. Moreover, PANA necessitates the existence of IP addresses in all the mesh nodes. This poses a serious constraint on deployment of this protocol.

Prasad et al. have presented a lightweight *authentication, authorization and accounting* (AAA) infrastructure for providing continuous, on-demand, end-to-end security in heterogeneous networks including WMNs [106]. The notion of a security manager is used by deploying an AAA broker. The broker acts as a settlement agent, providing security and a central point of contact for many service providers.

Lee et al. propose a distributed authentication scheme for minimizing authentication delay in a wireless network [107]. In this scheme, multiple trusted nodes are distributed over a WMN which act on the behalf of an authentication server. Deployment of multiple authentication servers makes management of the network easy, and it also involves less storage overhead in the MRs. However, the performance of the scheme will degrade when multiple MCs send out their authentication requests, since the number of trusted nodes acting as the authentication server is limited compared to the number of access routers.

The authentication schemes for MANETs can also be adapted in WMNs. Sen and Subramanyam have proposed and evaluated the performance of a distributed certificate authority based on threshold cryptography [108]. The scheme is an extension of the MOCA protocol [109] in which a collection of nodes selected on several parameters acts as the certificate authority and provides an attack resilient and robust certificate distribution and verification service.

In the following sub-sections, we provide a brief discussion on a few authentication schemes for WMNs. For a more comprehensive discussion on this topic, interested readers may refer to [3].

**3.6.1 ARSA: an attack-resilient security architecture for multihop WMNs**

Zhang and Fang have proposed an attack-resilient architecture for large-scale WMNs that deploys three categories of network entities: (i) brokers, (ii) users, and (iii) network operators [110]. In this architecture, each WMN domain is assumed to be operated by an operator, and it consists of a certain number of mesh routers. The mesh routers are assumed to be powerful in computing and communication capabilities. Hence, the packets transmitted by a mesh router reach their intended mesh client nodes (in the coverage area of the mesh router) in a single hop. On the other hand, the communication from the mesh clients to their mesh router may be multi-hop in nature because of the limited computing and transmission power of the mesh clients. Each user (i.e., mesh client) acquires a *universal pass* form the network operator which allows it to get ubiquitous access to the network. Multiple network operators need to have bilateral *service level agreements* (SLAs) between them. Instead, each network operator only needs to have an agreement with one or more brokers. The number of brokers is far less than the number of network operators since the WMN is very large in scale. For authentication purpose, the mesh clients need to locally communicate with its serving WMN domain without requiring any communication with the corresponding broker. This approach reduces authentication delay and signaling overhead in the authentication process. In addition, ARSA also provides efficient mechanisms for mutual authentication between any pair of node belonging to the same WMN domain. The scheme is not only efficient but also has been shown to be secure against various kinds of attacks such as: attack on location privacy, DoS attacks, bogus beacon flooding attack, bandwidth exhaustion attack etc [110].

ARSA assumes multiple trust domains in the mesh network, each domain being managed by a broker or a network operator. For accessing network services, each mesh client has to first register with at least one broker. Upon successful registration, the broker issues an *electronic universal pass* to the client. Each network operator also needs to establish trust relationship with one or more brokers. A network operator allows network access to a mesh client which has a valid universal pass issued by a broker with which the network operator has pre-established trust relationship. The passes are the important components in the authentication process in ARSA. In order to minimize the bandwidth and signaling overhead in the authentication process, the size of the passes is made as small as possible utilizing the concept of *identity-based cryptography* (IBC). Although the concept of IBC was first introduced by Shamir [39], a fully functional IBC scheme was not established till Boneh and Franklin applied *Weil pairing* to construct a *bilinear map* [40]. Use of IBC in ARSA requires the presence of one network entity in each trust domain. This entity, known as the *domain administrator*, performs some essential *trust domain initialization* activities [110].

ARSA uses three types of passes: (i) *router passes* which are issued by a network operator to its mesh routers, (ii) *client passes* which are provided by a broker to its registered clients, and (iii) *temporary client passes* which are issued by a network operator to the mesh clients present in its domain. ARSA utilizes router passes and client passes to realize authentication and key agreement between a mesh router and a mesh client and also between a pair of mesh clients. The authentication may be either inter-domain or intra-domain. In case of inert-domain authentication, a mesh client migrates from one WMN domain to another. In intra-domain scenario, a mesh client changes its association from its current mesh router to a new mesh router in the same domain. While inter-domain authentication is a more expensive operation than authentication in intra-domain, it occurs less frequently. ARSA provides a very efficient way for client-client authentication by using temporary client credentials for clients which are associated with the same mesh domain. In summary, the protocol provides an efficient, secure and ubiquitous network access to the users of a WMN and is ideally suited for large-scale networks with limited client mobility.

### 3.6.2 AKES: an efficient authenticated key establishment scheme for WMNs

He et al. have proposed a distributed *authenticated key establishment scheme* (AKES) that is based on *hierarchical multi-variable symmetric functions* (HMSF) [111]. In the proposed scheme, MCs and MRs can mutually authenticate each other and establish pair-wise communication keys without the need of any interaction with a central authentication server. This leads to reduced communication overhead and delay in the authentication process.

The WMN architecture assumed in the scheme is same as that used in ARSA scheme discussed in Section 3.6.1. The mesh network is divided into a number of *domains*. Each *mesh client* (MC) registers itself in its *home domain*. Each domain is managed by an *Internet service provider* (ISP) which relies on an *authentication authorization and accounting* (AAA) server for managing the entities in its domain. There are a few *Internet gateways* (IGWs) and *mesh routers* (MRs) and a large number of *mesh clients* (MCs) in a domain managed by an ISP. The scheme assumes pre-existing security frameworks for communications between AAAs and IGWS, between MRs and IGWs and between the MRs themselves. The goal of the scheme is to secure the communication sessions between the MCs and between the MRs and the MCs. In designing the proposed scheme, the authors have utilized the concept of *polynomial-based key generation* concept proposed by Blundo et al. [112]. Using the polynomial-based key distribution scheme an authorized server (e.g., AAA server) distributes a small piece of information (e.g., coefficients of polynomials) among a group of users in such a way that each user can compute a shared key with every other user in the group by only exchanging their IDs. These shared keys are used for pair-wise authentication and encryption of messages among the users. The proposed authentication scheme extends the concept of polynomial-based key generation so that it can be applied for *asymmetric mutual authentication* and key establishment. The need for asymmetric mutual authentication arises since the MRs and MCs in a WMN are unequal entities. Using the symmetric polynomial and asymmetric function, the authors have designed a "*hierarchical multi-variable symmetric function based authenticated key establishment scheme*" [111] that finally enables generating the MR-MC keys, MC-MC keys, and pair-wise session keys.

The robustness of the pair-wise keys is dependent on selection of the pair-wise master key generation functions [111]. In fact, it has been proved that for a polynomial of degree $t$, the polynomial-based key distribution scheme will be $t$-secure [113]. A $t$-secure key distribution scheme is robust against collusion attacks by a maximum of $t$ nodes. Therefore, the robustness of the keys can be increased by increasing the value of $t$ in the polynomial used for key generation. Since MCs authenticate locally with the MRs, spoofing attacks and DoS attacks in the wireless backbone involving the IGWs and

AAA servers are very difficult to launch. The computation overhead of the scheme depends on the function used in the key generation.

### 3.6.3 SLAB: a secure localized authentication and billing scheme for WMNs

Zhu et al. have proposed a *secure localized authentication and billing* (SLAB) scheme for wireless mesh networks to address security requirements and performance efficiency in terms of reducing inter-domain handoff authentication latency, and computation load on the roaming broker [114].

Most of the security solutions for WMNs are based on *authentication, authorization and accounting* (AAA) architecture [115], in which the authentication request from a *mobile user* (MU) is sent through the *serving mesh access point* (sMAP) and the *mesh gateway* (MGW) to a centralized authentication server (e.g., RADIUS server) that can grant access to the MU after verifying the authenticity of the MU. Since such lengthy authentication processes involve unacceptable delay in real-time applications, faster solutions are in demand. One approach that is followed for fast *inter-domain authentication* (authentication of an MU when it is roaming in a domain other than its home domain) is to have a pre-established trust among different *wireless Internet service providers* (WISPs) by deploying a centralized *roaming broker* (RB) trusted by all the WISPs [116]. In this approach, the foreign WISP, in whose domain the MU is currently roaming, forwards the current AAA session of the MU to the home WISP of the MU for authorization via the RB. For enhanced security, the RB may be configured not only as a *trusted third party* (TTP), but it also serves as a centralized *certificate authority* (CA) that issues public key certificates to the WISPs and MUs. This enables faster trust establishment among the WISPs or between a WISP and MUs since it only requires verification of the *public key certificates* (PKCs) issued by the RB [110, 117].

In designing the proposed scheme, the authors have observed that from the point of view of the WISPs, an inter-domain handoff can be considered as an inter-WISP payment, while from the point of view of the MUs, an MU can roam into another WISP domain only if it has enough credits remaining with it. From this perspective, a WISP can issue a digital signature based on PKI which is equivalent to a digital currency for inter-domain roaming payment with another WISP. This transaction does not involve any intervention of the RB. Moreover, this digital signature can also be taken as an authentication credential of the MU to which it is issued. The authors have called such digital signature as *D-coin* (digital coin).

The authors have also observed that if the *mesh access points* (MAPs) are pre-loaded with necessary cryptographic mechanisms, some important security-related operations - e.g., roaming/handoff authentication and billing -- can be performed in a localized manner with much better scalability and efficiency, thereby solving the scalability problem with respect to a centralized RB. However, this localized approach to authentication leads to security issues. The MAPs are low cost devices and susceptible to easy compromise [103]. An attacker can retrieve the cryptographic keys from a compromised MAP and launch some serious attacks such as *coin fraud attacks* (arbitrary issue of D-coins to an illegal MU or accepting D-coin from an MU and not providing service against it) [114].

SLAB exploits the advantages provided by localized authentication approach while providing adequate security protections against the vulnerabilities associated with the MAPs. To thwart coin fraud and overcharging attacks [114], a *local voting* strategy and *threshold digital signature* mechanism are adopted in designing the authentication scheme [118]. Local voting strategy enforces a requirement that the issued D-coin is not only endorsed by the *serving MAP* (sMAP) but also by the *neighboring MAPs* (nMAPs) to avoid a possible single-point-of-compromise at the sMAP. A *local user accounting profile* (LUAP) for the MU is maintained both at the sMAP and the nMAP for recording each network access and roaming operation performed by the MU. Since LUAPs for each MU are maintained, the on-line billing can be done easily. SLAB also enables inter-domain handoff authentication and billing to be done in a peer-to-peer manner without any intervention of the RB when an MU performs an inter-domain handoff. The RD is involved only during the "clearance phase" in which a WISP submits its collected d-coin issued by the other WISP for receiving payment against the D-coin. This operation is performed off-line and does not put much load on the RD. However, to further reduce the load on the RD during the off-line clearance phase, SLAB exploits the use of short and aggregate digital signature to minimize the overhead due to the verification and storage of the D-coin [119].

In summary, SLAB has five phases in its operation: (i) signing key distribution, (ii) secure session maintenance and LUAP generation phase, (iii) localized LUAP transfer during intra-domain handoff phase, (iv) D-coin issuance and inter-domain handoff authentication phase, and (v) clearance phase.

In the signing key distribution phase, for issuing a D-coin on behalf of the MGW, each of the MAPs first obtains its own share (i.e., partition) of the signing key using the threshold digital signature technique. In the secure session maintenance and LUAP generation phase, the sMAP of an MU collaborates with some of the nMAPs to generate and maintain the LUAP of the MU in order to track the spending information of the MU. In order to ensure the authenticity of the LUAP information exchange over a secure session, the MUs are mandatorily required to submit non-repudiation proof of the previous spending information so that the session consistency is maintained. The localized LUAP transfer during intra-domain handoff phase is involved in ensuring that every new nMAP of the MU can obtain a copy of the MU's authentic LUAP. During an intra-domain handoff, an MU roams within a common WISP resulting in a switch of the sMAP and the corresponding nMAPs. SLAP reduces signaling overhead during intra-domain handoff by invoking a localized LUAP transfer algorithm based on a local voting strategy. The algorithm accepts an LUAP as a valid one only if more than *k* valid LUAP copies from the nMAPs are found to be consistent. In the D-coin issuance and inter-domain phase, the MU roams in domains managed by different WISPs. Inter-domain handoff involves mutual authentication between the MU and the target WISP and the inter-domain WISP payment-related issues. SLAB handles inter-WISP authentication and billing using D-coin. In the clearance phase, the RB handles the inter-WISP payments in an efficient manner. An event-driven clearance procedure is used in which D-coin is regarded as an event. Since the processing is done in a batch mode, the D-coin can be only submitted to the RB when a given size of D-coin has been gathered or a specified time interval has elapsed. The batch processing enables the RB to verify the gathered D-coin and perform aggregate signature [119] simultaneously so that transmission and verification cost is minimized.

### 3.6.4 LHAP: a lightweight hop-by-hop authentication for ad hoc networks

Zhu et al. have proposed a *light-weight hop-by-hop access protocol* (LHAP) for authenticating data packets and preventing resource consumption attacks [70, 120]. The protocol uses a light-weight hop-by-hop authentication approach in which intermediate nodes authenticate the data packets they receive before forwarding them to the next hop node. LHAP employs one-way hash chains [121] for traffic authentication and the TESLA [71, 85, 122] protocol for bootstrapping and maintaining trust between the nodes.

A one-way hash chain is a chain of keys generated through repeated application of a one-way hash function on a random number. For example, if a node needs to generate a key chain of size $N$, it first chooses a key - $K(N)$ - randomly. The node, then, successively computes the remaining keys in reverse order so that $K(N-1) = F(K(N))$, $K(N-2) = F(K(N-1))$… till it get $K(0) = F(K(1))$. To use one-way hash chain for the purpose of authentication, a sender node first signs the last value in the chain (i.e., $K(0)$) with its private key so that another node that has the knowledge of its public key can verify the signature and the authenticity of $K(0)$. The sender, then, discloses the successive keys in the chain in the reverse order in which they were generated. The receiver can verify the authenticity of $K(j)$ by checking $K(j-1) = F(K(j))$.

LHAP uses the TESLA protocol [71, 85, 122]. TESLA is a broadcast authentication scheme that uses a one-way hash chain and an approach of delayed key disclosure. After bootstrapping an authentic key derived from a one-way hash chain, the sender digitally signs it and sends it to the receivers. TESLA computes the *message authentication codes* (MACs) in the subsequent broadcast authentications but discloses the key at the receivers with a delay. In the basic scheme of TESLA, the sender node uses a key $K$ from its hash chain as the MAC key to compute a MAC over packet $P(i)$, and then appends the MAC to $P(i)$. The key $K$ is disclosed in the next packet $P(i + 1)$, which allows the receiver nodes to verify the authenticity of the key $K$ and hence the MAC of $P(i)$. If both $K$ and the MAC are verified to be correct, and if the packet $P(i)$ is guaranteed to be received before the packet $P(i + 1)$ was sent, the receiver nodes conclude that the packet $P(i)$ is authentic. A critical requirement of TESLA is a receiver's ability to determine the sending time of each packet. This requirement is met by periodic key disclosure and loose time synchronization [85].

LHAP assumes that each node in the network has a public key certificate signed by a trusted *certificate authority* (CA), and the public key of the CA is known to all the nodes in the network. A loose time synchronization scheme is also assumed to be available so that the TESLA protocol can be used.

Since LHAP authenticates all traffic packets at each node on the route from the source to the destination, it is mandatory that the authentication protocol should be lightweight. In the LHAP data packet authentication, each source node generates a one-way hash chain of keys which are used by its immediate neighbor nodes. The keys generated from this one-way hash chain are termed as "TRAFFIC KEYs". Each neighbor of a node obtains an authentic key in this TRAFFIC KEY chain when it first establishes a trust relationship with the node. A node transmitting a packet will append a new TRAFFIC KEY to the packet. All the neighboring nodes that receive this packet can verify the its authenticity by checking the validity of the attached TRAFFIC KEY.

One naïve way of bootstrapping trust among the node is to exchange TRAFFIC KEYs using public key encryption in which each node signs its most recently released TRAFFIC KEY and sends it to each of its neighbors. However, this approach lacks scalability in a large-scale dense wireless network. To solve this problem, LHAP uses TESLA to minimize the number of signature operations. EACH node only uses digital signatures to bootstrap a TESLA key chain, and the TESLA keys are used to generate subsequent authentic TRAFFIC keys. To maintain the trust relationship, each node broadcasts its latest TRAFFIC KEY at periodic intervals. The broadcast TRAFFIC KEYs are authenticated using the TESLA keys of the node. After the updated TRAFFIC KEY from a node is received, all of its neighbors drop any packets they receive which are authenticated by the older TRAFFIC key. This periodic broadcast of KEYUPDATE messages ensures that the protocol is secure against any possible replay attack.

Security analysis of LHAP shows that the protocol is robust against various outsider and insider attacks [70, 120]. Among the outsider attacks, the protocol is resistant to attacks such as: impersonation attack, wormhole attack, hidden terminal attack etc. LHAP is also resistant to insider attacks such as: insider-clone attack. However, it cannot defend against multiple insider attacks launched by multiple insider compromised nodes.

### 3.6.5 A localized two-factor authentication scheme for WMNs

Lin et al. have proposed a two-factor localized authentication scheme for inter-domain handover and mobility management in IEEE 802.11 standard compliant WMNs [123]. The localized authentication scheme is based on Rabin cryptosystem [124]. Rabin cryptosystem has asymmetric computational overheads – while the encryption and signature verification operations are very fast, the decryption and signature generation operations are computationally intensive [125].This asymmetric property of Rabin cryptography makes it particularly suitable for scenarios like inter-domain handover in wireless networks where the *access points* (APs) have high computational power and the *mobile stations* (MSs) have limited resources.

The authentication scheme is designed for a standard wireless network depicted in Fig.11 that consists of four types of entities: (i) the *mobile units* (MUs), (ii) a *trusted third party* (TTP), (iii) the *wireless Internet service providers* (WISPs), and (iv) the *hotspots*. As shown in Fig.11, several hotspots may be under the operation of a single WISP, and these hotspots may not necessarily be adjacent to each other. Each hotspot is assumed to have one AP. The use of the terms AP and hotspot can be interchangeable. Before an MU can access the network services, it has to first subscribe to the TTP. The TTP and each of the WISPs have mutual agreement such that an MU after subscribing to the TTP can access the hotpot operated by the corresponding WISP. The TTP also acts as the *certificate authority* (CA) that issues certificates to the MUs and WISPs. The issued certificates are essentially the public keys of the MUs and the WISPs which are digitally signed by the TTP. After successful registration, each MU is also issued a *smart card* that contains the authentication credentials of the MU. The smart card serves as an electronic pass for the MU for roaming across different WISPs.

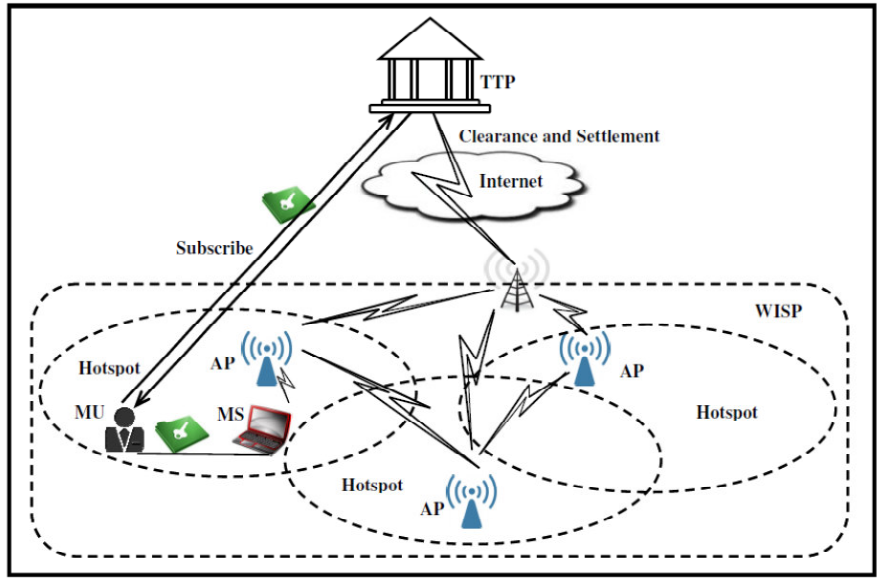

**Fig 11. WMN architecture and different network entities in the local authentication protocol [122]**

The authentication protocol has three phases: (i) the *initialization* phase, (ii) *the login and mutual authentication* phase, and (iii) the *handoff* phase.

In the initialization phase, various operations are performed by different entities in the network. The TTP generates its public-private key pair and publishes its public key while keeping the private key secret. Each WISP sets up a mutual agreement with the TTP such that a registered and authenticated MU can log in and access the network services provided by the WISP. For this purpose, each WISP sends a chosen identity of it to the TTP and also generates its private key. The TTP checks the validity and uniqueness of the identity. If the validation process passes correctly, TTP signs the public key of the WISP and also sets up a linkage between the public key and the identity of the WISP. The TTP sends this signed public key of the WISP and the linkage of the public key with the identity of the WISP to the concerned WISP. After the initialization of the WISP is complete, its starts the initialization process for the hotspots operated by it. Each hotspot generates its own private key and its *service set identifier* (SSID). The hotspot sends its SSID to its home WISP for verification and keeps its private key secret. The home WISP checks for uniqueness of the SSID. If the SSID provided by the hotspot is found to be unique, the home WISP create a signature on the public key of the hotspot and also sets up a linkage between the public key of the hotspot and its SSID. The home WISP sends the signed public key of the hotspot and the linkage of the public key and the SSID to the concerned hotspot thereby completing the initialization of the hotspot.

The next step of the initialization process involves registration of the MUs. For this purpose each MU generates a unique identity of it and sends the identity to the TTP through a secure channel. After checking for the uniqueness of the identity of the MU, the TTP provides a smart card to the MU which can be used as an electronic pass by the MU.

To provide enhanced security, the scheme uses two-factor authentication for a roaming MU. In order to defend against scenarios where the confidential security credentials of an MU might have been leaked, the second factor using a smart card is adopted. If an attacker has to successfully impersonate a legitimate user, the attacker must not only know the authentication credentials of the user, but also capture or replicate the corresponding smart card. Two-factor authentication provides increased robustness to the localized authentication approach.

In the login and mutual authentication phase, the MU and the hotspot (i.e., the AP) mutually authenticate each other using a robust challenge and response mechanism [123].

The handoff procedure is performed between an MU and APs when the MU moves from the current AP to an adjacent AP. In order to minimize the delay in handoff, the proposed localized authentication scheme adopts an approach in which the session key is cached in the current network domain. When an MU sends a handoff request to an AP with which the MU shares a valid shared session key, an authenticated symmetric key agreement protocol is invoked instead of a full authentication procedure that involves asymmetric key cryptographic operations. This makes the handoff operation fast while compromising on the security in the authentication process.

The security analysis of the localized authentication scheme has shown that it is resistant to replay attack, impersonation attack, password guessing attack, and attack on the privacy of the users (i.e. MUs).

Table 5 provides a brief summary of a few authentication mechanisms for WMNs.

**3.7 Security mechanisms against attacks on key management protocols**

Several key management protocols for wireless networks exist in the literature. Due to the computational resource constraints in the mobile hand-held devices, schemes which involve high computational overhead for security-related operations are not particularly suitable for wireless networks. Again, the Diffie-Hellman key exchange protocol [32] and many of its variations do not fit to the requirement of an online *certificate authority* (CA) that computes the shared keys between the nodes on-demand in a WMN. In this section, we provide a brief overview of a few key management protocols in wireless networks which can be adapted to WMNs. For a detailed discussion on these schemes, the readers may refer to [177].

Capkun et al. have proposed a *public key infrastructure* (PKI)-based key management approach for wireless multi-hop networks [126]. The proposed scheme does not use any trusted third party or a CA for key distribution and management purposes. Instead, it follows an approach of using self-signed certificates as in *pretty good privacy* (PGP) [127]. Based on their individual experiences, the users (i.e., nodes) issue public key certificates to each other. Each node maintains a repository of certificates that it has issued to others and that have been issued to it by other nodes. The certificate revocation process can be explicit or implicit. In explicit revocation process, the issuer of the certificate explicitly informs other nodes about the certificates being revoked. In the implicit revocation approach, the nodes use expiry time for each certificate. The certificates which have not been renewed before the expiry of their time of validity are assumed to be revoked. In this scheme, if node *A* wants to communicate to node *B*, the two nodes first merge their certificate repositories before node *A* tries to find a public key certificate chain that has a path from node *A* to node *B*. The chain is constructed progressively in such a way that the first certificate can be verified using node *A*'s public key, and each of the subsequent certificates can be verified using the key included in the previous certificate in the chain. The last certificate must include the public key of the node *B* so that the verification process is successfully completed. The authors have introduced trust-based confidence metrics which are included in each certificate so as to defend against any possible attacks by malicious nodes that may issue false certificates.

Table 5. Summary of some authentication schemes for WMN communication

| Protocol | Salient Features |
|---|---|
| **ARSA [110]** | This efficient authentication architecture assumes a WMN to be consisting of multiple trust domains, each domain being managed by a broker. Each client registers with a broker which issues an electronic universal pass to the client. The pass serves as the authenticated credential. ARSA supports inter-domain and intra-domain mobility of the clients efficiently by providing efficient and ubiquitous network access to the clients. It is resistant to attacks on location privacy of the users, DoS attacks, bogus beacon flooding attack and bandwidth exhaustion attack. |
| **AKES [111]** | It is a distributed authenticated key establishment scheme for WMNs. The WMN consists of a number of domains each having an AAA server for managing the entities. A hierarchical multi-variable symmetric function-based authenticated key establishment scheme is used for generating the MR-MC keys, MC-MC keys, and the pair-wise session keys. The robustness of the pair-wise keys generated depends on the selection of the pair-wise master key generation functions. Due to the localized authentication used by the MCs with the MRs, spoofing attacks and DoS attacks in the wireless backbone network is very difficult to launch on this scheme. The computational overhead depends on the polynomial function used in the key generation. |
| **SLAB [114]** | This authentication scheme exploits the advantages provided by localized authentication approach while providing adequate security protections against the vulnerabilities associated with the mesh access points (MAPs). The scheme works in five phases: (i) distribution of keys, (ii) secure session maintenance and local user accounting profile (LUAP) generation phase, (iii) localized LUAP transfer during intra-domain handoff phase, (iv) D-coin issuance and inter-domain handoff authentication phase, and (v) clearance phase. SLAB provides a strong authentication framework while supporting user mobility by reducing the delay in handoff. |
| **LHAP [70, 120]** | It is a light-weight hop-by-hop access control and authentication protocol for data packets that is resistant to resource consumption attack. It uses one-way hash chains for data traffic authentication and the TESLA protocol for bootstrapping and maintaining trust between the nodes. It assumes that each node has a public key certificate signed by a trusted CA, and the public key of the CA is known to all the nodes. A loose time synchronization scheme is also assumed to be available to support TESLA. The protocol is robust against various outsider and insider attacks such as: impersonation, wormhole, hidden terminal, insider-clone attacks. However, it is vulnerable to cooperative insider attacks launched by multiple insider compromised nodes. |
| **Localized two-Factor Authentication [123]** | This authentication scheme supports inter-domain handover and mobility in IEEE 802.11 compliant WMNs. It utilizes the asymmetric property of Rabin cryptosystem in mobility management issues where APs have high computational power and the MSs are resource-constrained. For providing enhanced security, it uses two-factor authentication for roaming mobile users. The scheme is resistant to attacks such as: replay, impersonation, password guessing and attack on the privacy of the users. |

Based on the concept of *self-certified key* (SCK) [128], Li and Garcia-Luna-Aceves have proposed a protocol named NIKAP for facilitating key agreement processes in WMNs [129]. In the NIKAP protocol, pair-wise keys are computed between two nodes in a non-interactive manner. The services of a CA are needed only during the initial phase of the network bootstrapping. The non-interactive progression capability of NIKAP makes it particularly suitable for WMNs in which a pair of nodes needs to establish pair-wise shared keys without any negotiation over an insecure channel.

Zapata has proposed a key management system named *simple ad hoc key management* (SAKM) that allows the nodes of an ad hoc network to use asymmetric key cryptography for providing security and robustness to multi-hop routing used in such networks [130].

Key distribution protocols using ID-based cryptography [40] or the combination of threshold and ID-based cryptography [131] have the same advantage as SCK because IDs are used to obtain the corresponding public keys of nodes, instead of using a certificate to bind the ID and its public key. However, online CA services must exist for such protocols to work. The requirement of an online CA service is the drawback of an ID-based cryptosystem and a threshold cryptography-based system [132].

An approach to combine threshold secret sharing and probabilistic key sharing is proposed by Zhu et al. [133]. In this scheme, a source node splits its shared secret key with the destination node into several parts and sends the parts to the destination node in such way that the destination node can reconstruct the full key from the individual parts with a high probability. However, in addition to having high overhead of communication (due to transmission of several parts of the key) and intensive computation overhead (in splitting and recovering the shared secret key), the protocol fails in its operation if at least a threshold minimum number of parts do not reach the destination node.

In group key agreement protocols [134, 135], usually a shared key is distributed among all the members of a multicast group. While these protocols have low storage complexity compared to the schemes that use pair-wise keys for each node pair in a network, the cost of re-keying operations in these protocols is very high because of leaving and joining of nodes in the multicast group. Moreover, if a group key is compromised, it will lead to a complete collapse of security in the group as opposed to a possible compromise of a pair-wise secret key which will affect only the concerned node pair.

In order to increase the efficiency and robustness in the certificate construction process by combining the secret shares in a threshold cryptography-based system, Joshi et al. propose to allocate each designated distributed CA node more than one share of the CA's secret [136].

Signature aggregation [137] mechanism aggregates all the certificates in a chain of certificates into a single short signature so as to reduce the size of the certificate chain. The resultant smaller certificate incurs less communication overhead and consumes less bandwidth. The basic principle of signature aggregation is based on the idea that if $N$ distinct messages are signed by $N$ distinct signing authorities (i.e. nodes), then it is possible for a trusted node in the network to aggregate all these $N$ signatures into a single signature in a secure way such that a verifier node at the receiver side can securely verify that each user indeed signed its own message.

**3.8 Security mechanisms against attacks on privacy protection schemes**

The issue of user privacy in WMNs has attracted serious attention from the research community. Several propositions have been made in the literature for protecting privacy of user data, location and other sensitive information in wireless communication networks.

Wu et al. propose a light-weight privacy preserving solution to achieve a well-maintained balance between network performance and traffic privacy preservation [5]. At the center of the scheme is an information-theoretic metric called *traffic entropy*, which quantifies the amount of information required to describe the traffic pattern and to characterize the performance of traffic privacy preservation. The authors have also presented a penalty-based shortest path routing algorithm that maximally preserves the traffic privacy by minimizing the mutual information of traffic entropy observed at each individual relaying node while controlling the possible degradation of network within an acceptable region. Extensive simulation results demonstrate the effectiveness of the solution and its resilience to a possible collusion between two malicious nodes. However, one of the major problems of the solution is that the algorithm is evaluated in a single-radio, single channel WMN. The performance of the algorithm in multiple radios, multiple channels scenario will be a really questionable issue. Moreover, the solution has a scalability problem.

Wu and Li have proposed a mechanism with the objective of hiding an active node that connects to a gateway router, where the active mesh node has to be anonymous [6]. A communication protocol is designed to protect the node's privacy using both cryptography and redundancy. The protocol uses the concept of *onion routing* [138]. A mobile user who requires anonymous communication sends a request to an *onion router* (OR). The OR acts as a proxy to the mobile user and constructs an onion route consisting of other ORs using the public keys of the routers. The onion is constructed in such a way that the inner most part is the message for the intended destination and the message is wrapped by encrypting it using the public keys of the ORs in the route. The mechanism protects the routing information from insider and outsider attack. However, it has a high computation and communication overhead.

Sen proposes a reputation-based trust management system has been combined with a privacy preservation scheme for designing a secure and efficient searching protocol for unstructured and decentralized peer-to-peer networks [139]. The scheme utilizes network topology adaptation by constructing an overlay of trusted peers where the nodes in a neighborhood are selected based on their trust ratings. The overlays of trusted peers are utilized to achieve a user-preserving searching protocol.

To control the misuse of personal information and to prohibit disclosure of personal data, different types of information hiding mechanisms like anonymity, data masking can be implemented in WMN applications. The following approaches can be useful in information hiding, depending on what is needed to be protected:

- *Anonymity*: this is concerned with hiding the identity of the sender or receiver of the message or both of them. In fact, hiding the identity of both the sender and the receiver of the message can assure communication privacy. Thus, the possible attackers who may monitor the messages being communicated could not know who is communicating with whom.
- *Confidentiality*: it is concerned with hiding the transferred messages by using suitable data encryption algorithms. Instead of hiding the identity of the sender and the receiver of a message, the message itself is hidden in this approach.
- *Use of pseudonyms*: this is concerned with replacing the identity of the sender and the receiver of the message by pseudonyms which function as identifiers. The pseudonyms can be used as a reference to the communicating parties without infringing on their privacy. This ensures that the users in the WMNs cannot be traced or identified by malicious adversaries. However, it is important to ensure that there is no indirect way by which the adversaries can link the pseudonyms with their corresponding real world entities.

In the following sub-sections, we provide a brief discussion on some well-known scheme for privacy protection in WMNs.

### 3.8.1. Traffic privacy preservation using penalty-based multipath routing

Privacy is a critical issue in the context of WMN-based Internet access by users in their residences where users' traffic is forwarded via multiple mesh routers [5]. Since the inbound and outbound traffic of a residence can be easily observed by the mesh routers residing at the neighboring locations, there is no privacy protection of the users. In order to provide protection to traffic privacy of the users in a community mesh network, Wu et al., propose a privacy-preservation scheme which is resistant to any traffic analysis of the users' sensitive personal communications [5]. The scheme exploits the intrinsic redundancies in WMNs by routing traffic from a source node to a destination node through *multiple paths*. The traffic is split in a random manner both in spatial and temporal domains so that the intermediate nodes can only have limited knowledge of the traffic pattern. This *traffic concealment* mechanism makes traffic analysis an impossible proposition. More formally, in this scheme, the data packets are routed in such a way that the statistical distributions of the traffic data as observed in each intermediate node on the routing path are *independent* of the actual traffic data between the source-destination pair. An information theoretic metric -*traffic entropy*- is defined that quantitatively identifies the amount of information required for profiling a traffic pattern. Based on the computation of traffic entropy, a penalty-based routing is proposed that minimizes the mutual information of traffic entropy observed at each intermediate node on the multiple routing paths.

The scheme assumes a three-tier architecture of WMN, in which the client devices communicate with stationary mesh routers. Multiple mesh routers communicate with each other to form a wireless backbone and a group of mesh routers are connected to an Internet gateway by wired links (or high-speed wireless links). Each mesh node and the gateway node have a public-private key pair. The gateway maintains a directory of certified public keys of all mesh nodes. Each mesh node knows the public key of the gateway. Before the start of each session, a pair of mesh nodes first establishes a secret key (derived using the public keys of the node pairs and the public key of the gateway). The shared secret key is used to encrypt all messages in that session. When an IP packet is sent from a source node ($s$) in the Internet to a client ($d$) in the mesh network, the packet is encrypted at the gateway using the shared key between the gateway ($g$) and the client ($d$). If $i$ is the mesh router of the client $d$, to route the encrypted packet to destination node $d$, the gateway $g$ prefixes to the packet the source route from the gateway $g$ to the mesh router $i$. The encapsulated packet is then forwarded by relaying routers in the WMN. Since the source address and all higher layer information including the port numbers are encrypted, the intermediate routers cannot access any information about the source with which the client is communicating. However, this encryption scheme can only provide data confidentiality and it is easy to make traffic analysis on the scheme. Since the source route is transmitted in clear text form in an encapsulated packet, the mesh routers can observe the traffic pattern of a mesh client node. To address this problem, traffic from the gateway to the mesh routers is routed through multiple paths in such a way that for each individual packet the chosen route provides traffic information to routers in the route that is independent of the overall traffic.

The penalty-based routing algorithm for traffic privacy preservation executes in three phases: (i) *path pool generation*, (ii) *candidate path selection*, and (iii) *individual packet routing*.

In the *path pool generation phase*, a large number of paths from the gateway node ($g$) to the destination node ($x$) are identified. The paths are identified in such a way that the majority of them are disjoint paths, i.e., they do not have any common nodes. The set of these paths is denoted as $S_{paths}$. Each node is assigned a penalty weight. The weight of a link is the average of the penalty weights at its two end nodes. The weight (i.e. cost) of a path is the sum of the penalty weights of all the edges that constitute the path. The path pool generation algorithm works iteratively. At the beginning of the first iteration, each node is assigned a penalty weight of 1. The *Dijkstra's shortest path algorithm* [140] is utilized to identify the shortest path between the gateway ($g$) and the destination node ($x$). Once, the first shortest path is identified, the penalty weights associated with each nodes on this route are increased arbitrarily so that these nodes do not become potential candidates in the next round of execution of the algorithm. The algorithm is executed iteratively to find the optimum paths in successive iterations. The iteration continues till the set $S_{paths}$ has acquired sufficient number of candidate paths.

In the *candidate path selection phase*, a subset $S_{selected}$ is chosen from the set $S_{paths}$. The elements of $S_{selected}$ are chosen randomly from the set $S_{selected}$. After a path is selected in the set $S_{selected}$, to reduce its probability of getting selected again in the next round, a suitable factor is employed. The use of this probability factor prevents selection of multiple identical paths in the set $S_{selected}$.

In the packet routing phase, for each packet, one path from $S_{selected}$ is randomly chosen and the packet is routed through that path. Every time a particular path is chosen for routing a packet, the value of a counter corresponding to that chosen path is increased by one. If the value of this counter for a path reaches a threshold value, then the entire $S_{selected}$ set is assumed to have expired and a new $S_{selected}$ is chosen by invoking the *candidate path selection phase* once again. Since each packet is routed through a randomly selected path, and the candidate paths are mostly disjoint, the probability that the packets are routed through the same path will be negligibly small. The algorithm seeks to achieve a tradeoff between *routing efficiency* (i.e. routing through the shortest path between a source-destination pair) and *traffic pattern privacy* (routing through disjoint multipaths).

The authors have also shown that the proposed mechanism is robust against collusion attack by two neighboring nodes which exchange their knowledge (*colluded traffic mutual information*) about the same destination. However, the traffic splitting increases delay in communication and hence this mechanism lacks scalability. It may not, therefore, be suitable for large-scale WMN deployment scenarios delivering real-time applications.

### 3.8.2. PEACE: Privacy-enhanced yet accountable security framework

Ren et al. have proposed a novel privacy and security scheme named PEACE (Privacy-Enhanced yet Accountable security framework) for WMNs [141]. The authors have assumed a three-tier architecture of a WMN under the control of a *network operator* (NO). At the top level, there are gateways connected to the public Internet, at the next level there are *access points* (APs) and *mesh routers* (MRs) which are connected to the Internet gateways by either high-speed wireless links or high-speed wired link. The MRs and APs are also interconnected among them by high-speed wireless links forming a WMN. At the lowest level of the architecture, there are wireless mobile devices which are connected to the MRs and APs by lower-speed wireless links. The security associations between the gateways and the MRs and APs are assumed to be already in place. The goal of PEACE is to secure the communication between the mobile devices with the MRs and APs. In view of this, the security objectives of PEACE are to achieve the following: (i) authentication and key management between the user devices and the routers, (ii) mutual authentication and key agreements among the users (i.e., mobile devices), (iii) user privacy protection, (iv) user accountability, and (v) membership maintenance.

To protect user privacy, one essential requirement is that the user information should be well protected from network communications against an adversary and even a network operator (NO). To satisfy this requirements, PEACE aims to achieve the following: (i) no communication sessions should reveal any information related to the identity of a user except that the user is a legitimate member in the network, (ii) no adversary and NO should be able to link two different communication sessions to a particular specific user in the network, and (iii) a given communication session under the audit by an NO can only be linked to the attributes of the user without disclosing his/her full identity, (iv) each user is assumed to belong to a user group and each user group is supposed to be managed by a group manager. The full identity of a user can be identified only by the joint effort from the user group manager and the NO and not by any of them individually.

To address the aforementioned security and privacy requirements, PEACE proposes a trust and key management model. The high-level trust model includes four network entities: (i) the network operator, (ii) user group managers, (iii) user groups, and (iv) a *trusted third party* (TTP). Each user group consists of a set of users according to certain aspects of their *non-essential attributes*. Each group has one group manager who is delegated with the responsibility of adding or removing users from that group. The manger of a group knows the essential and non-essential attributes of each of the members in his group since these information is shared by the users to the manager at time of their joining the network. Each group manager subscribes to the network operator on behalf of its group members. The network operator allocates a set of *group secret keys* to the user group after successful registration of the group. Each group secret key is divided into two parts before they are distributed by the network operator. One part of the key is sent to the corresponding group manager and the other part is communicated to the TTP. To access network services, each user sends two requests: one request to its group manager and the other to the TTP for recovering the complete group secret key. On receiving the key parts, the user sends signed acknowledgments to the group manager and the TTP. This principle of key management which is based on *separation of powers* enables PEACE to achieve a high level of robustness in its security and privacy protection strength. From the network operator's point of view, access security is guaranteed since each user has to gather a valid complete group secret key and generate a valid access credential based on that key. User privacy is highly enhanced since the user identity information and the corresponding key information are divided among three autonomous network entities – the group manager, the TTP and the network operator. Although the network operator has the access to the complete user secret key, it does not have any information about how these keys are mapped to the essential attributes of the users. On the other hand, the group manager and the TTP both know the essential attributes of the users, but none of them have complete information about the secret key. Unless two among the three entities (network operator, TTP and group manager) collude, PEACE makes it impossible for any of them to have access to a user's essential attributes from the access credentials submitted by the user. However, the accountability of a user is fully ensured since in case of any possible incident, a legal authority can collect information from the network operator, the group manager, and the TTP to precisely identify the user.

Since the conventional blind signatures and ring signature schemes can only provide irrevocable anonymity, and PEACE requires revocable anonymity to guarantee user accountability, the authors have developed a variation of the short group signature scheme proposed in [142] to meet the requirement. PEACE also provides a robust mechanism for dynamic addition and deletion of users and mesh routers to and from the network. A comprehensive security analysis has shown that the scheme is resistant to bogus data injection attacks, data phishing attacks and DoS attacks [141].

### 3.8.3 SAT: A security architecture for achieving anonymity and traceability

Sun et al. present a security architecture named "SAT" [143, 144]. The system uses ticket-based protocols for resolving the conflicting security requirements of unconditional anonymity for honest users and traceability of misbehaving users in a WMN. By using the tickets, self-generated pseudonyms, and the hierarchical identity-based cryptography, the architecture has been demonstrated to achieve the desired security objectives while maintaining the performance efficiency at a high level. The system uses a blind signature technique from payment systems [145-148] to achieve anonymity by delinking user identities from their activities. The pseudonym technique also renders user location information unexposed. The pseudonym generation mechanism does not rely on a central authority unlike the *broker* in ARSA [110], the *domain authority* in [149], the *transportation authority* or the *manufacturer* in [150], and the *trusted authority* in [151], who can derive the user's identity from his/her pseudonyms and illegally trace on an honest user. However, the system is not intended for achieving routing anonymity. *Hierarchical identity-based cryptography* (HIBC) for inter-domain authentication is adopted to avoid domain parameter certification in order to ensure anonymous access control.

Fig. 12 depicts the architecture of a WMN on which the SAT scheme is implemented. The scheme assumes a wireless mesh backbone consisting of the *mesh routers* (MRs) and the *gateways* (GWs) which are connected by wireless links. The MRs and GWs serve as the access points for individual WMN domains such as hospital, campus etc. It is assumed that each domain is managed by a *trusted authority* (TA). Each TA is connected to its GWs by high-speed links. The TAs and the GWs are powerful computing devices and these devices are physically protected from any possible attacks. The scheme assumes the existence of a mechanism that distributes pre-shared keys and establishes secure communication channels between the TAs, the GWs, and the MRs at the backbone. SAT, however, focuses on the authentication and the key establishment problems during the network access of the MCs. It adopts the *hierarchical ID-based signature scheme* (HIDS) [152] for inter-domain authentication of clients which are affiliated to a home TA and are currently visiting a foreign TA. For trust domain initialization, the authors have proposed the use of hierarchical *identity based cryptography* (IBC) [152]. SAT uses a *ticket-based security architecture* for client authentication, data integrity, and confidentiality in communications both in the home domain and in inter-domain. The ticket-based architecture is responsible for ticket issuance, ticket deposit, fraud detection, and ticket revocation protocols. When a client joins the network, the TA issues a ticket to it. The client needs to reveal its real ID to the TA at this time so that TA can verify the authenticity of the client.

The client employs some blinding techniques to transform the ticket so that it cannot be linked to a specific execution of the ticket generation algorithm while the verifiability of the ticket is maintained. This ensures that the TA cannot link the issued ticket to the real identity of the client. The authors have proposed use of a *partially restrictive blind signature scheme* [153, 154] in the ticket generation algorithm. The TA publishes the domain parameters to be used by the clients in its trust domain following the standard IBC domain initialization process. After obtaining a ticket, a client may deposit the ticket anytime for accessing network services before the ticket expires. The ticket is to be deposited only once to the first encountered gateway that provides the network access. The gateway (also called the *deposit gateway*) verifies the signature and the integrity of the deposited ticket and generates a signature on the client's pseudonym along with the ID of the *deposit gateway* (DGW). This signature is required so that the other access points in the trust domain can determine whether and where to forward the client's access request, in case the deposited ticket is to be further used from other access points. The client is not allowed to change its pseudonym, since the DGW will refuse network access to the client if the current pseudonym does not match with the one that was initially recorded with the ticket. The real ID of a client has to be revealed to the home TA at the time the ticket is issued due to the requirement of client authorization. However, if the privacy of the client is to be protected, this ID is hidden from the access point. Unless the access point colludes with the home TA, user identity privacy cannot be compromised. For obtaining new tickets, the client simply deposits the old ticket using the ticket deposit protocol and gets the new ticket. The request for ticket is sent to the home TA in encrypted form.

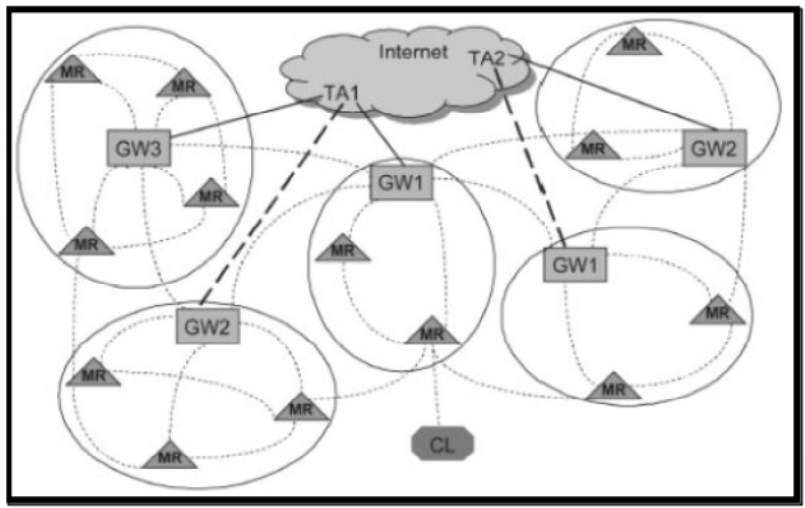

Fig 12. The WMN architecture used in SAT scheme [144]

The ticket revocation protocol is used for two purposes: (i) revocation of new tickets and (ii) revocation of deposited tickets. The client may have some unused tickets which have not been deposited before. For revoking these tickets, the client sends a signed and time-stamped request for revocation to a GW. The GW authenticates the client and records the ticket serial number as revoked. For revoking already deposited tickets, the client sends an authenticated revocation request to the DGW. The DGW verifies the authenticity of the client and marks the associated ticket as revoked. A GW after successfully executing a revocation request, reports the revocation details to the home TA. The home TA updates and distributes the revocation list for all GWs in the trust domain.

When a client attempts to access network services from a foreign (visiting) trust domain, two possible scenarios of authentication for the client may arise. In the first case, the client does not have any available ticket for accessing the network from the foreign domain. In this case, a foreign *mesh router* (MR) forwards the client's new ticket request to the home domain of the client. In the second scenario, the client has available new tickets issued by its home TA. In this case also the foreign mesh router forwards the tickets to the home domain for verifying the authenticity of the available tickets.

In SAT, the clients use a robust and efficient way for generating their pseudonyms. The method of self-generation of pseudonym is similar to that proposed by Rahman et al. in [155]. Self-generation of pseudonyms by clients incurs very low overhead and hence, the clients can use it frequently to regenerate their pseudonyms. This increases the strength of the anonymity and privacy protection.

### 3.8.4. An efficient and user privacy preserving routing protocol for WMNs

Sen has proposed an efficient and secure routing protocol for WMNs for handling stringent *quality of service* (QoS) requirements of real-time applications while ensuring protection of user privacy [156]. The scheme assumes a three-tier architecture of WMNs in which the top level contains the *Internet gateways* (IGWs) which are connected to the wired Internet forming the backbone of the WMN. At the next level of the architecture are the wireless mesh routers which are connected to the IGWs and among each other forming a wireless mesh. The mesh clients (i.e., the wireless mobile user devices) are at the lowest level of the architecture which can connect to the MRs using wireless networking for accessing the Internet services. The author has identified several challenges in designing an efficient routing protocol such as: (i) accurate measurement of link reliabilities, (ii) robust estimation of the end-to-end delay in routing, (iii) reducing control overhead to maximize throughput in the network. To address these challenges the proposed protocol has incorporated several salient features such as: (i) designing a robust estimate for reliability of each routing path for a source-destination pair, (ii) exploiting the network topological information for efficient route discovery, (iii) developing a reliable model for estimating the end-to-end delay along a routing path, (iv) use of *multi-point relay* (MPR) nodes to reduce the control overhead in routing, (v) accurately estimating the available bandwidth in a wireless link and along a wireless path for a source-destination pair, (vi) routing through a path in the fixed network through wired IGWs, (vii) identification of selfish nodes, (viii) protecting the user privacy. In the following, these features are described very briefly. A more detailed discussion can be found in [156].

In the protocol, every node computes the reliability of each of its out-bound wireless links. For computing the reliability of a link, the number of control packets a node receives during a given time interval is used as the base parameter. An *exponentially weighted moving average* (EWMA) approach is use to update the link reliability estimate for a link [156]. Each node maintains and periodically updates a *link reliability table* after computing the reliability metric for each of its out-bound wireless links. The reliability for an end-to-end routing path is computed based on the average of the reliabilities of all the links constituting the path.

The protocol also exploits the knowledge of network topology by using a strategy of selective flooding of control messages, and hence avoids broadcast of control messages. If a source-destination pair is under the control of the same MR, the flooding of the control message are made only within the part of the network served by the MR. To further reduce the overhead of control messages, the nodes accept broadcast control messages from only those neighbors which have link reliabilities greater than a threshold minimum value. The paths with less reliability values are not discovered and hence not considered for routing.

For achieving an accurate estimation in end-to-end delay over a routing path, the protocol uses *probe packets* during the route discovery phase. In this approach, when a source node receives the RREP packet from a destination in response to an RREQ message, it constructs a table and stores all the RREP packets that have been received along with the routes through which the RREP messages have arrived. The source node then sends some probe packets to the destination through the (reverse) path along which the first RREP message arrived from the destination. This ensures that the probe packets are sent along the path which is likely to induce the minimum end-to-end delay. The number of probe packets is kept limited to $2H$ for a path of $H$ hops to make a tradeoff between the control overhead and the measurement accuracy. The destination node computes the average delay experienced by all the probe packets it has received, and sends the computed value to the source node piggybacking it on an RREP message. If the computed average delay is less than the maximum delay that can be tolerated to satisfy the application QoS, the source node selects the route. Otherwise, the source tries with the next path in its table till it finds a path that satisfies the delay tolerance.

For further reducing the control overhead, the protocol selects MPR nodes as in *optimized link sate routing* (OLSR) protocol, and routes the control packets through these nodes. In addition to the computation of path reliability and use of MPR nodes to minimize the overhead due to flooding, the protocol uses a robust estimation of the available bandwidth of the wireless paths used in routing by computing the packet loss due to congestion in the wireless links [156].

The efficiency in routing is further improved by occasional routing of packets through the fixed wired network backbone. Every MC knows its hop distance from its IGW. This hop count information is included in the RREQ messages from the MC. When the destination MC receives the RREQ, since it also knows its hop distance from its gateway, it checks which path is better from routing – the path through the WMN or the path through the fixed network. If the destination node finds that the better route is through the fixed network, it sends the RREP message through wired backbone network. Therefore, in such situations, the forward route is established between the source and the destination through the wired network, while the reverse route is setup through the WMN. Since the nodes in the forward and the reverse routes are on node-disjoint paths, they do not contend for accessing the wireless medium resulting in an improved throughput and reduced end-to-end delay. For handling any possible selfish behavior of nodes (a selfish node utilizes network resources for routing its own packet but avoids forwarding packets for other nodes to conserve its energy), the protocol employs a simple mechanism to discourage selfish behavior and encourage cooperation among nodes. Each node forwards routing packets to only to those neighbors which have link reliability greater than a threshold value. Since the link reliabilities for the selfish nodes are all zero (the selfish nodes do not forward any packets), these nodes never get any opportunity to participate in routing. Hence, packets originating from these nodes are always dropped. The link reliability metric serves the dual purpose of enhancing the reliability and enforcing cooperation among the nodes.

The protocol also ensures user anonymity and privacy protection by using a *novel* approach that extends the ring signature authentication scheme [157]. A security analysis of the protocol has shown that the protocol ensures: (i) *user anonymity*, (ii) *anonymous authentication of a user with an authentication server (AS) and also anonymous authentication among a pair of users*, (iii) *forward secrecy* (forward secrecy of a scheme refers to its ability to defend against leaking of its keys of the previous sessions when an attacker is able to catch hold of the key of a particular session). Since forward secrecy is ensured, the protocol is resistant to any replay attack [156].

Table 6 presents a summary of some privacy protection schemes in WMN communication.

### 3.9 Cross-layer security mechanisms

Kidston et al. have proposed a cross-layer architecture for security in multi-hop wireless networks [158]. The authors argue that by utilizing the metrics from the security services at one layer, such as authentication systems and *intrusion detection systems* (IDS), robustness in the security services being provided at other layers can be increased. For example, application layer security services such as secure authentication and IDSs can provide real-time attack profiles into an integrated cross-layer security service. These profiles may be used by the lower layers to improve their detection efficiencies since the lower layers don't have to compute any security metrics at all. While the computational complexity in a node increases for using this cross-layer security services, the communication overhead in the network is reduced to a large extent. In the proposed cross-layer architecture, the authors have used a *publish-subscribe* system that acts a vertical interface between all the layers of the communication protocol stack [158]. This interface is used as a real-time messaging framework by the layers for communicating critical security metrics. Other layers subscribe to the metrics. An independent cross-layer service may also be designed that will exclusively subscribe to the security metrics. Each protocol stack focuses only on its core security functionality while subscribing to the value-added metrics published by other layers. The per-layer information is combined to build a consolidated security picture that can assist the operator of the network in designing an automated response system for responding to any possible attack.

Table 6. Summary of some privacy preservation schemes in WMN communication

| Protocol | Salient Features |
|---|---|
| **Penalty-based multipath routing [5]** | This mechanism aims to achieve a balance between network performance and traffic privacy in a WMN. It utilizes a novel concept called traffic entropy that quantifies the amount of information required to describe the traffic pattern and to characterize the performance of traffic privacy preservation. The scheme uses a penalty-based shortest path routing algorithm that maximally preserves the traffic privacy by minimizing the mutual information of traffic entropy observed at intermediate node while ensuring that network performance is maintained at high level. The scheme is effective in preserving traffic privacy and resilient against colluding malicious nodes. However, it has a scalability problem. |
| **PEACE [141]** | It is a privacy architecture for WMNs that assumes three-tiers with the gateways connected to the Internet at the topmost level, MRs at the second level and user mobile devices at the lowest level. The security associations between the gateways and the MRs are assumed to be in place. The protocol has the following objectives: (i) ensure authentication and robust key management between the user devices and the MRs, (ii) provide mutual authentication and key distribution among the mobile user devices, (iii) protect the privacy of the users, (iv) ensure user accountability in the system, and (v) maintain the node membership in the system. The protocol has a robust trust and key management system to satisfy its security and privacy requirements. In addition to providing revocable anonymity, the protocol is resistant to bogus data injection attack, data phishing attacks and DoS attacks. |
| **SAT [143, 144]** | The protocol uses tickets, self-generated pseudonyms, and a hierarchical identity-based cryptographic structure to achieve the conflicting tradeoff between security and user privacy while maintaining the performance efficiency at a high level. The pseudonym generation technique does not rely on a central authority and provides user location privacy while avoiding any single point of failure in the system architecture. The protocol provides anonymous access control to the users but cannot ensure routing anonymity. The ticket-based security architecture provides client authentication, data integrity, and confidentiality in communications both in the home domain and in inter-domain. A blinding technique is used by a client to transform the issued ticket by the TA (trusted authority) so that the TA cannot link the issued ticket to the real identity of the client. |
| **Privacy preserving routing [156]** | It is a secure routing protocol that can handle stringent quality of service (QoS) requirements of real-time applications while ensuring protection of user privacy. Various modules of the protocol perform the following functions: (i) accurate measurements of wireless link reliabilities, (ii) robust estimation of end-to-end delay in routing, (iii) reducing the control packet overhead and maximizing the throughput in the network, (iv) use of multi-point relay (MPR) nodes to reduce control overhead, (v) accurate estimation of the available bandwidth in the wireless links on the routing path for a source-destination pair, (vi) identification of selfish nodes, (vii) protection of user privacy. User anonymity and privacy is ensured by a novel scheme that extends the ring signature authentication scheme. |
| **Privacy in P2P networks [139]** | This protocol is designed to combat the problem of unauthentic file downloads as well as to improve the search scalability and efficiency in an unstructured peer-to-peer network while protecting the privacy of the users. The adaptive trust-aware protocol is based on the construction of an overlay of trusted peers where neighbors are selected based on their trust ratings and content similarities. The semantic communities of peers also enable the protocol for form a neighborhood of trust that is utilized to protect user privacy. |
| **Anonymity using onion routing [6]** | The objective of this mechanism is to hide an active node that connects to a gateway router for accessing the Internet, where the active mesh nodes are to be kept anonymous. The communication protocol uses the concept of onion routing (OR) to protect the privacy of the node. The mobile user who requires anonymous communication sends a request to the onion router which constructs an onion route consisting of the public keys of the routers. The onion is constructed in such a way that the inner most part is the core message which is wrapped by encrypting it using the public keys of the ORs on the route. The mechanism, however, has a high computing and communication overhead. |

## 4. Trust Management in WMNs

The use of trust and reputation-based frameworks for enforcing security and collaboration among nodes in wireless multi-hop networks is a very popular approach. A trust-based scheme can protect against attacks that are sometime beyond the capabilities of cryptographic mechanisms. For example, issues like judging the quality and reliability of wireless links for establishing routing paths in high-throughput multicast routing, robust key management, reliable packet forwarding over multiple hops etc. can be elegantly addressed by utilizing the services of a trust-based framework. Several trust- and reputation-based frameworks for mobile ad hoc and wireless sensor networks exist in the literature [8]. Most of these schemes can be easily adapted to WMNs.

Pirzada and McDonald propose an approach to building trust among the nodes in an ad hoc network that can be deployed in a WMN which does not have a centralized trusted entity [159]. In this scheme, the nodes passively monitor the packets received and forwarded by their neighbors. The receiving and forwarding of packets are considered as events. The events are assigned different weights based on the applications and the forwarding behaviors of the nodes. The weights reflect the significance of the concerned event with respect to the associated applications. The trust values associated with all the events of a node are combined to arrive at an aggregate trust metric for the node. The computed trust values of a pair of nodes are used for deriving the trust value of the link connecting those nodes. The links which have higher trusted values are assigned smaller weights and a shortest-path algorithm is utilized to find the most trusted path (i.e., the path with the minimum weight) between a pair of nodes.

Yen et al. propose a security solution for ad hoc networks that is based on a trust framework to ensure data protection, secure routing, and message authentication [160]. In this scheme, logical and computational trust analysis and evaluation methodologies are applied in each node. Each node then evaluates the trust metric for each of its peers using factors such as experience statistics, data value, intrusion detection results, and recommendations from its neighbors leading to a robust trust management framework.

Sen et al. have proposed a trust establishment mechanism among nodes in a MANET [161]. The proposed framework is a probabilistic solution based on a distributed trust model. A *secret dealer* is introduced during the network bootstrapping phase to initiate the trust establishment process. Subsequently, the nodes use a self-organized certificate distribution and management procedure based on threshold cryptography. The mechanism has been shown to be very effective in a large-scale dynamic ad hoc network.

In a multi-operator WMN deployment, the major issue is the lack of trust between the heterogeneous network entities belonging to different service providers and network operators. In such scenarios, reputation-based systems are useful for establishing and sustaining trust between mobile users and different network operators and service providers. Ben Salem et al. have addressed the issue of interoperability for trust building in a wireless network [162]. The authors argue that since the home network of a service provider can be the home network for some mobile nodes and a foreign network for some other mobile nodes, the home network of a service provider cannot always be trusted by all mobile nodes. The proposed reputation-based system allows mobile nodes to evaluate the trustworthiness of a service provider based on the latter's behavior. At the same time, the system also allows a service provider to authorize mobile users to access its network services based on their trust levels with the service provider.

Jarrett and Ward have presented a trusted computing-based routing protocol- *trusted computing ad hoc on-demand distance vector* (TCAODV) [163] – that extends the base *ad hoc on-demand distance vector* (AODV) [83] routing protocol. In TCAODV, each node has its public key certified by a *certificate authority* (CA) and the public key certificate of each node is stored within a trusted root. The source node broadcasts its public key certificate with its *HELLO* message. Each of the neighbor nodes, on receiving the certificate, first verifies the authenticity of the certificate and then stores it as the broadcaster's public key. The RREQ packet sent by each node is signed using the integrity metric from the routing module of the sender. Each neighbor, on receiving the RREQ message, verifies the signature using the previously stored broadcaster's key and determines whether the provided integrity measurements are trustworthy. The intermediate nodes on the path from the source to the destination strip off the signature in the RREQ message and put their own signatures and integrity measurements. A *per-route symmetric encryption key* is also established for allowing only the trusted nodes along the route to use the route. All traffic sent along this route is encrypted using this symmetric key.

Sen presents a reputation- and trust-based security framework for *mobile ad hoc networks* (MANETs) that can detect packet dropping attacks by malicious insider nodes [164]. The mechanism uses a trust model that computes the reputation metrics for all nodes in the network based on their packet forwarding statistics. The proposition includes a robust trust metric computation, propagation and update process which involves low computation and communication overheads. Moreover, it can be easily adapted to WMNs which do not have any centralized point of administration or a *certificate authority* (CA) for key management or a deployed *trusted third party* (TTP).

Tang and Wu propose a trust-delegation-based *efficient mobile authentication scheme* (EMAS) that uses the elliptic curve discrete logarithm problem [165]. Lee et al. have presented a distributed authentication model for reducing the authentication delay by deploying multiple trusted nodes which serve the role of an *authentication server* in a WMN [107]. Zhang et al. propose an *attack resilient security architecture* (ARSA) for WMNs that aims at providing secure roaming in multi-domain WMNs using a *user-broker-operator trust model* [110]. Lin et al. have presented an authentication scheme for WMNs that utilizes the services of a *trusted third party* (TTP) which also acts as a trusted *certificate authority* (CA) for issuing certificates to service providers and mobile users [123].

For establishing a trust relationship among different *wireless Internet service providers* (WISPs) in a multi-operator WMN, an approach of deploying a centralized *roaming broker* (RB) has been proposed by Leu et al. [116]. The RB is trusted by all the participating WISPs. In this approach, when a *mobile user* (MU) roams into the domain of a foreign network, the foreign WISP forwards the AAA session of the MU to its home WISP for authorization verification via the RB. In the scheme proposed by Long et al., the RB not only acts as TTP, but also as a certificate authority that issues public key certificates to the WISPs and the MUs [117]. The trust relationships among the WISPs or between a WISP and MUs can be easily established by validating the public key certificates by the RB. The foreign WISP reports the accounting information of the roaming MU to its home WISP at the completion of the session. Although, the deployment of an RB effectively reduces handoff authentication delay, unfortunately, it becomes a point of performance bottleneck and single-point-of-failure. Zhu et al. have presented a digital signature-based inter-domain roaming and billing architecture wherein a WISP not only accepts a valid public key certificate issued by another WISP but also accepts the digital signature issued by that WISP as a payment technique [166]. Hence, inter-domain authentication and billing are performed simultaneously by verifying and validating a digital signature. The scheme assumes that every *access point* (AP) is enough trustworthy is authorized to issue digital signatures on behalf of its WISP. However, this assumption may not be realistic in large-scale distributed WMNs in which some of the *mesh access points* (MAP) may be easily compromised.

Several other trust and reputation-based schemes are available for wireless multi-hop networks such as: Watchdog and Pathrater [59], CONFIDANT [167], CORE [168], RFSN [169], DRBTS [170], OCEAN [171]. The interested readers may refer to [8] for further details on these schemes.

**5. Security in WMNs – Open Problems and Future Research Challenges**

WMNs have become the focus of research in wireless networks in the recent years owing to their great promise in realizing numerous next-generation wireless services. Driven by the demand for rich and high-speed content access with stringent QoS requirements, recent research on WMNs has focused on developing high performance communication protocols, while the security and privacy issues of these protocols have received relatively less attention. However, given the wireless and multi-hop nature of communication in WMNs, these networks are vulnerable to a wide variety of attacks at all layers of the communication protocol stack. Although, the researchers have made substantial contributions in the areas of security and privacy in WMNs, there are still many challenges that remain to be addressed. Some of these challenges and a few future research directions in security for WMNs are briefly discussed in the following.

Existing authentication schemes for WMNs under high mobility conditions involve large delay and latency that sometimes adversely affect the QoS of the applications and network services. Efficient, lightweight and robust authentication protocols for MRs and MCs with high mobility need to be designed that minimize authentication and handoff delay. For this purpose, scalable key management systems are also required. Similarly, for reliable routing in wireless environment that involves high rate of packet drops, energy-aware, efficient multipath routing schemes are also in demand for high throughput applications.

For strategic deployments of WMNs, robust hop integrity verification protocols need to be designed. In addition, due to hardware/software compatibility and efficiency considerations, it may be a challenging issue to design a strategic plan for deploying the hop integrity verification protocols in the nodes in a WMN. While the deployment of an integrity verification protocol in each MR may be a very secure strategy, from efficiency point of view this may not be an optimal one. An adaptable strategy that can be varied based on the requirements of the applications will be ideal.

Existing security schemes for deployment in hybrid WMNs are also inadequate. In hybrid WMNs, the networks are designed to be integrated with other types of networks, such as wired networks and cellular wireless networks. Such networks are vulnerable to a wide range of attacks which are not present in the individual network components. For instance, a mesh network for wireless Internet access can be targeted with DoS attacks launched from the Internet [172]. The scarcity of bandwidth resource on the wireless network makes the attack more difficult to defend and the existing defense mechanisms for DoS attacks will be ineffective in this scenario. On the other hand, hybrid networks also present additional resources and opportunities for defending against attacks. For example, for defending attacks on a WMN connected to a wired network, it is possible to leverage the high bandwidth, low latency wired links, and powerful computers in the wired network for handling security issues in the wireless networks.

Authentication and key management in WMNs with highly mobile MCs and MRs pose particularly difficult challenges. Owing to the very limited coverage of IEEE 802.11-based MRs (e.g., 100 meters), high mobility users migrate very fast from the coverage area of an MR to another. It is not acceptable for a user to authenticate and negotiate the key with each MR. Novel solutions possibly by using group keys are needed for this purpose. A very interesting research direction in this regard is to investigate whether an approach of *signature aggregation* [137] can be utilized for developing an efficient key management scheme for WMNs. This can be extremely useful for group key establishment leading to reduction in the overhead incurred in the group key creation and re-keying process.

Use of novel approaches like *dynamic topology-aware adaptations* and *dynamic network coding* for securing multicast messages in WMNs are very interesting emerging trends [172]. The essential principle of dynamic topology adaptation is to improve network performance by dynamically adjusting the protocol structures based on the variations of the wireless links. On the other hand, dynamic network coding exploits the broadcast nature of the wireless medium and makes use of the common occurrence of packet overhearing in wireless networks with the ultimate goal to improve network performance. Ahlswede et al. have proposed the use of network coding for secure routing in wireless networks [173]. However, the existing network coding systems are vulnerable to a wide range of attacks besides the most well-known *packet pollution attacks* [174]. Many of the weaknesses of the existing systems lie in their single focus on performance optimizations. A more balanced approach, which can provide improved security guarantees, is crucial for the actual adoption of network coding in real-world applications. A future direction of research is to uncover the security implications of different design and optimization techniques, and explore balanced system designs with network coding that achieve appropriate tradeoffs between security and performance that is suitable for different applications' requirements.

Cross-layer security approach is another possible future direction of research for wireless networks. While cross-layer protocols for wireless networks have been widely researched in the literature [175], their applications in security domain have found very limited attention. As mentioned in Section 3.9, cross-layer approach to security could lead to efficient and proactive intrusion and anomaly detection in some mission-critical wireless network deployments.

For user privacy preservation, optimized deployments of emerging techniques like *fully homomorphic encryption* [176] in WMNs can be another very interesting direction of research.

## 6. Conclusion

WMNs have become an important focus area of research in the recent years owing to their great potentials in realizing numerous next-generation wireless services with stringent QoS guarantees and with high mobility support for the users. Driven by the increasing demand for rich, high-speed and bandwidth intensive content access, recent research has focused on developing high performance communication protocols for such networks, while issues like security, privacy, access control, intrusion detection, secure authentication etc. have taken the back seat. However, given the inherent vulnerabilities of the wireless medium due to its broadcast nature and multi-hop communications in WMNs, these networks are subject to a wide range of threats. This chapter has made a comprehensive presentation on the various attacks on different layers of the communication protocol stack of WMNs. While highlighting various vulnerabilities in the physical, link, network, transport and application layers, this chapter has also focused its attention on how attacks can be launched on authentication, privacy and key management protocols on WMNs. After identifying various security threats, the chapter has presented a comprehensive state of the art survey on various defense mechanisms for defending those attacks. Some of these defense mechanisms are also compared with respect to their different approaches towards security and their performance efficiencies. Finally, some of the emerging trends in research and future research issues related to security and privacy in WMNs are presented.